\documentclass[12pt,preprint]{aastex}
\usepackage{natbib}
\begin{document}

%\title{Keck Interferometer Observations of T Tauri Stars:
%Inner Disk Structure, Magnetospheric Accretion, and Planet Formation}
%\title{Keck Interferometer Observations of T Tauri Disks at Sub-AU Radii:
%Magnetospheric Accretion and Planet Formation}
\title{Observations of T Tauri Disks at Sub-AU Radii:
Implications for Magnetospheric Accretion and Planet Formation}

\author{J.A. Eisner\altaffilmark{1}, L.A. Hillenbrand\altaffilmark{1},
R.J. White\altaffilmark{1}, R.L. Akeson\altaffilmark{2}, 
\& A.I. Sargent\altaffilmark{1}}
\email{jae@astro.caltech.edu}
\altaffiltext{1}{California Institute of Technology,
Department of Astronomy MC 105-24,
Pasadena, CA 91125}
\altaffiltext{2}{California Institute of Technology, 
Michelson Science Center MC 100-22,
Pasadena, CA 91125}


\slugcomment{Draft of {\bf \today}}

\begin{abstract}
We determine inner disk sizes and temperatures for four solar-type (1-2
M$_{\odot}$) classical T Tauri stars (AS 207A, V2508 Oph, AS 205A, and
PX Vul) using 2.2 $\mu$m observations from the 
Keck Interferometer. Nearly contemporaneous near-IR adaptive optics imaging
photometry,  optical photometry, and high-dispersion optical spectroscopy 
are used to distinguish contributions from the inner disks and
central stars in the interferometric observations.  In addition, the
spectroscopic and photometric data provide estimates of  stellar properties,
mass accretion rates, and disk co-rotation radii.  We model our 
interferometric and photometric data
in the context of geometrically flat accretion disk models with inner holes, 
and flared disks with puffed-up inner walls.
Models incorporating puffed-up inner disk walls generally provide better
fits to the data, similar to previous results for higher-mass Herbig Ae stars.
Our measured inner disk sizes are larger than disk
truncation radii predicted by magnetospheric accretion models, with
larger discrepancies for sources with higher mass accretion rates.  We suggest
that our measured sizes correspond to dust sublimation radii, and that 
optically-thin gaseous material may extend further inward to 
the magnetospheric truncation radii.  Finally, our inner disk measurements
constrain the location of terrestrial planet formation as well as
potential mechanisms for halting giant planet migration.
\end{abstract}

\section{Introduction \label{sec:intro}}
T Tauri stars are low-mass ($\la 2$ M$_{\odot}$) pre-main sequence objects,
thought to be early analogs of stars like our own
Sun.  A wealth of evidence, including direct imaging at millimeter and 
optical wavelengths \citep[e.g.,][]{KS95,DUTREY+96,MO96}, and modeling
of spectral energy distributions \citep[SEDs; e.g.,][]{ALS88,BBB88,
BECKWITH+90}, has confirmed 
the long-espoused hypothesis that T Tauri stars are surrounded by massive
disks of dust and gas.  Moreover, observed line profiles and UV continuum
excesses indicate that T Tauri stars are actively 
accreting material from their circumstellar disks \citep[e.g.,][]
{WALKER72,EDWARDS+94,GULLBRING+98}.

The structure of the innermost disk regions 
may reveal the mechanism by which material is accreted
through the disk onto the star.  
In the current paradigm, T Tauri disks are 
truncated by the stellar magnetosphere
within the co-rotation radius, with material accreting along magnetic
field lines onto high-latitude regions of the star 
\citep[e.g.,][]{KONIGL91,SHU+94}.  For typical T Tauri star masses,
radii, magnetic field strengths, 
and accretion rates, predicted truncation radii
range from $\sim 0.02-0.2$ AU.
Observational measurements of these truncation radii are an obviously 
important test of the theory of magnetospheric accretion.

The spatial and temperature structures of
inner circumstellar disks are also important for understanding the 
properties of dust and gas in the terrestrial planet region,
and ultimately for understanding the formation of planets.
For example, a puffed-up inner disk wall, 
due to the normal angle of incidence of 
stellar radiation on the truncated inner edge \citep[e.g.,][]{DDN01},
may lead to shadowing, and thus cooler temperatures in the inner disk
compared to standard flat or flared disk temperature profiles 
\citep[e.g.,][]{CG97}. This, in turn, would have profound implications as to 
how and where terrestrial planets form \citep[e.g.,][]{HAYASHI81,SL00}.
Furthermore, inner disk structure is important for understanding how
the close-in extra-solar planets discovered by radial velocity surveys
\citep[e.g.,][]{MB00} either formed at, or migrated to their observed
orbital radii \citep[e.g.,][]{LBR96}.

Currently, only near-IR interferometric observations have sufficient
spatial resolution to probe directly the geometry and temperature of 
hot ($\sim 1000-2000$ K) disk regions within $\sim 1$ AU of young stars. 
Observations of the inner disks of a few of the brightest T 
Tauri stars \citep{AKESON+00,COLAVITA+03} and of their more massive 
counterparts, Herbig Ae stars \citep{MILLAN+99,MST01,EISNER+03,EISNER+04},
demonstrated that inner disks around lower-mass stars ($\la 5$ M$_\odot$) 
are larger than inferred by fitting geometrically thin accretion disk
models to SEDs \citep[e.g.,][]{BECKWITH+90}.
Inclusion of puffed-up inner walls in the models leads to consistent
fits to both interferometric and SED data for these objects
\citep{EISNER+04,MUZEROLLE+03}.  On the other hand,
higher-mass Herbig Be stars \citep{EISNER+04} and the extreme
accretor FU Ori \citep{MALBET+98,MALBET+05} are fitted well with
simple flat disk models, suggesting that inner
disk structure may depend on accretion rates or stellar properties.
A larger sample of resolved inner disks, including lower-mass T Tauri stars,
is necessary to explore such trends.

Here, we present 2.2 $\mu$m
Keck Interferometer observations of the inner disks around 
four solar-type (1-2 M$_\odot$) T Tauri stars, 
potential analogs to our own young Sun.  In order to model 
the stellar and circumstellar emission accurately, we combine our
spatially resolved interferometry data with optical/near-IR SEDs
and high-resolution echelle spectra.  The photometric and spectroscopic
data are essential for decomposition of the observed 2.2 $\mu$m flux into 
stellar and excess components.
Since T Tauri stars are variable at near-IR wavelengths on timescales
of several days to months \citep[e.g.,][]{SKRUTSKIE+96}, 
our spectroscopic, photometric, and interferometric data were obtained
within several days of one another.  
The photometry and spectra also enable determination of various properties 
of these systems, including stellar masses, ages, temperatures, radii, 
$v \sin i$, binarity, mass accretion rates, magnetospheric truncation radii,
and co-rotation radii.  

From the 2.2 $\mu$m interferometry data, 
we establish inner disk radii and temperatures, and distinguish between flat
and puffed-up inner disk models. In addition, we compare these measured sizes 
with inferred magnetospheric and co-rotation radii.  Although our
sample is small, the range of stellar and accretion properties allows us
to explore how inner disk structure depends on these parameters.

\section{Observations and Data Reduction \label{sec:obs}}
\subsection{Sample}
Our sample consists of four classical T Tauri stars:
AS 207A, V2508 Oph, AS 205A, and PX Vul.
AS 207A, the optically-brightest T Tauri star in $\rho$ Oph, was
identified as a young star based on its H$\alpha$ emission \citep{SR49}, and 
also as 
one component of a $0\rlap{.}''6$ binary system \citep{GNM93}.
The T Tauri star V2508 Oph \citep{WALTER86}
is located near the L162 dark cloud.  
AS 205A is a well-known young star near the $\nu$ Sco dark nebula
\citep[e.g.,][]{MB50}, identified as the brightest component of a
$1\rlap{.}''3$ binary system by \citet{GNM93}.
We assume that AS 207A, V2508 Oph, and AS 205A 
are all at the approximate distance of the $\rho$ Oph cloud, 160 pc
\citep{CHINI81}.  Finally, PX Vul is a T Tauri star in the Vul R2 region,
at a distance of 420 pc
\citep[e.g.,][]{HK72,HERBST+82}.
Properties of our sample, including celestial coordinates, distances, and 
spectral types, are included in Table \ref{tab:sample}.
%, and photometric
%magnitudes at optical and near-IR wavelengths are listed in 
%Table \ref{tab:phot}. 
%Our sample has $V$ magnitudes ranging from 11.5 to 13.5 and $K$ magnitudes
%from 6.4 to 7.7.  

\subsection{2.2 $\mu$m Interferometry \label{sec:ki}}
We observed AS 207A, V2508 Oph, AS 205A, and PX Vul
with the Keck Interferometer (KI)
on June 2, 2004.  KI is a fringe-tracking long baseline near-IR Michelson
interferometer combining light from the two 10m Keck apertures 
\citep{CW03,COLAVITA+03}.  The fringe-tracker detects a source in
a 5ms integration, setting a limiting $K$-band magnitude of 
$m_K \sim 9$.  In addition, sources must be optically bright enough for the
adaptive optics (AO) systems on each Keck aperture. Superb seeing
($\la 0\rlap{.}''5$) allowed excellent AO performance for our sample.
% requires sources optically brighter than $m_V \sim 12$.  
%Due to superb observing conditions, we were able to extend this limit to 
%$m_V = 13.5$.  
%Our targets all satisfy these brightness constraints.

For each target, we measured squared visibilities ($V^2$) at K-band
($\lambda_0=2.2$ $\mu$m, $\Delta \lambda = 0.4$ $\mu$m).  The system
visibility (i.e., the point source response of the interferometer),
was measured using observations of unresolved calibrators,
weighted by the internal scatter in the calibrator and the temporal 
and angular proximity to the target source \citep{BODEN+98}. 
Source and calibrator data were corrected for 
detection biases as described by \citet{COLAVITA99} and averaged into 5s 
blocks. The calibrated $V^2$ for the target sources are the average of the 5s 
blocks in each integration, with uncertainties given by the quadrature addition
of the internal scatter and the uncertainty in the system visibility. 
Typical uncertainties are $\sim 5\%$.

All calibrators were chosen to be compact (angular diameters $\la 0.2$ mas)
and close to the target sources (within $\sim 10^{\circ}$).  
In addition, our calibrators have $K$-band
magnitudes similar to those of our targets, to minimize potential biases.  
At optical wavelengths, the 
calibrators are brighter than the targets, which may lead to
enhanced AO performance; by
measuring the photon counts along both interferometer arms and applying
a standard ``ratio correction'' \citep[e.g.,][]{COLAVITA99}, 
we calibrate out the effects of 
AO performance on the visibilities.  The data for AS 207A,
V2508 Oph, and AS 205A were calibrated using HD 142943 and HD 148968, and
data for PX Vul were calibrated using HD 181383 and HD 182919.

\subsection{$JHK$ Adaptive Optics Imaging \label{sec:ao}}
We obtained dithered imaging observations of our sources 
at $J$,$H$, and $K$ on June 4, 2004, 
using the Palomar 200-inch adaptive optics system \citep{TROY+00} with the 
PHARO camera \citep{HAYWARD+01}.  After bias correction, background 
subtraction, and flat-fielding of the images, photometric fluxes were measured
with respect to the same calibrators used in our KI observations.
Calibrator magnitudes are known from the 2MASS catalog, assuming they 
are non-variable.  The photometric
errors are given by the quadrature addition of the RMS variations in brightness
between source integrations and the uncertainties in the calibrator
magnitudes.  Since the sources and calibrators were observed at 
similar airmasses, we apply no atmospheric extinction corrections.
The measured fluxes for these sources are 
listed in Table \ref{tab:phot}.

With the high angular resolution afforded by adaptive optics imaging 
($\sim 0\rlap{.}''1$ at $K$-band) we were
able to resolve AS 207 and AS 205 into binaries, finding parameters (see
Table \ref{tab:binaries}) consistent with previous measurements 
\citep{GNM93,RZ93,KORESKO02,BKM03}.
For these systems, we measured photometric fluxes for
both the primaries and secondaries; $JHK$ magnitudes for the primaries are
listed in Table \ref{tab:phot}, and $\Delta JHK$ values are given in
Table \ref{tab:binaries}.  Since the projected binary separations are much
larger than the field of view of KI (50 mas),  we 
obtained interferometric data only for the primaries.
No spatially resolved companions brighter than $\Delta K = 5$ 
were detected near V2508 Oph or PX Vul, and we consider these to be single
stars hereafter.

\subsection{$UBVRI$ Photometry \label{sec:optical}}
We observed our sample through Johnson
$U$,$B$,$V$, and Kron $R$ and $I$ filters on June 8, 2004 
using the robotic Palomar 60-inch telescope. 
Photometric fluxes were measured from bias-corrected, flat-fielded images
using well-studied photometric standards \citep{LANDOLT92}.
We determined extinction corrections and magnitude zero-points
using observations of five Landolt standards obtained throughout
the night.  Photometric errors for our target sources are the sum of various
uncertainties in quadrature: the RMS variation between integrations
(where multiple integrations of a source are available), the 
uncertainties in zero-points and extinction coefficients, and
uncertainties in magnitudes of our calibrators.  Photometric
uncertainties are typically $\la 10\%$, except at $U$-band, where
substantial uncertainties in the extinction coefficients lead to large error 
bars for the measured fluxes.

The seeing-limited resolution of these observations was $\sim 1\rlap{.}''7$,
and the close binaries in our sample, AS 207 and AS 205, are
unresolved.  Optical photometry for these sources, listed
in Table \ref{tab:phot}, therefore 
includes contributions from both the primaries and the secondaries.

\subsection{High Resolution Optical Spectroscopy \label{sec:highres}}
High dispersion optical spectra of the sample were obtained on
2004 June 11 using the HIRES spectrograph \citep{VOGT+94} on Keck I.
HIRES was used with the red collimator, an RG-610 filter, and the D1 decker
(1\farcs15 x 14\farcs0), yielding R$\approx$ 34,000 spectra over 6330-8750
\AA, with gaps between orders.  An internal Quartz lamp 
and a ThAr lamp were observed with the same setup for flat fielding and 
dispersion correction.  Several dwarf spectral type standards with known
radial velocities were also observed to assist in the spectroscopic
analysis.  The binary AS 205 was observed with the slit along the axis of the 
pair, while AS 207 (an unresolved binary in these observations), V2508 Oph and 
PX Vul were observed with the slit
perpendicular to the horizon (vertical mode).

\begin{figure}
\plotone{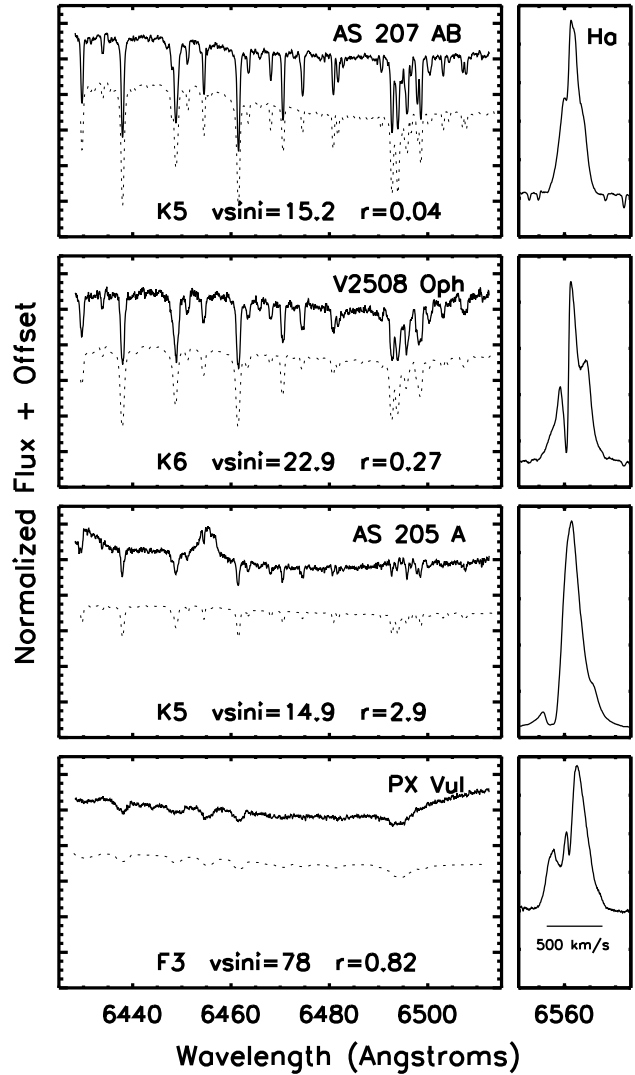}
\caption{Portions of the Keck/HIRES spectra within the R-band (left), and
H$\alpha$ emission profiles (right); both panels have the same 
wavelength scale.  
The best fit dwarf standards, rotationally
broadened and optically veiled, are shown as dashed lines for comparison.
The strong, broad H$\alpha$ emission profiles suggest all stars are
accreting.
\label{fig:spec}}
\end{figure}

The HIRES spectra were reduced using the facility ``makee'' reduction
script written by Tom Barlow.  Reduction includes bias-correction,
flat-fielding, spectral extraction,
sky subtraction, wavelength calibration, and heliocentric radial velocity
corrections.  This procedure worked well for single stars, but not for the
components of AS 205 whose spectra overlap.  In that case, the component
spectra were determined by fitting two Gaussians to each one-dimensional
cut in the spatial direction of the two-dimensional spectra.  The FWHM of
the best fit Gaussians were 0\farcs4; the seeing was quite good and the
1\farcs2 pair is reasonably well resolved.  These extracted component
spectra were then assigned the wavelength solution of the combined system
as determined by makee.  Portions of the extracted spectra for our 
sources are shown in Figure \ref{fig:spec}.

\section{Analysis \label{sec:anal}}

\begin{figure}
\plottwo{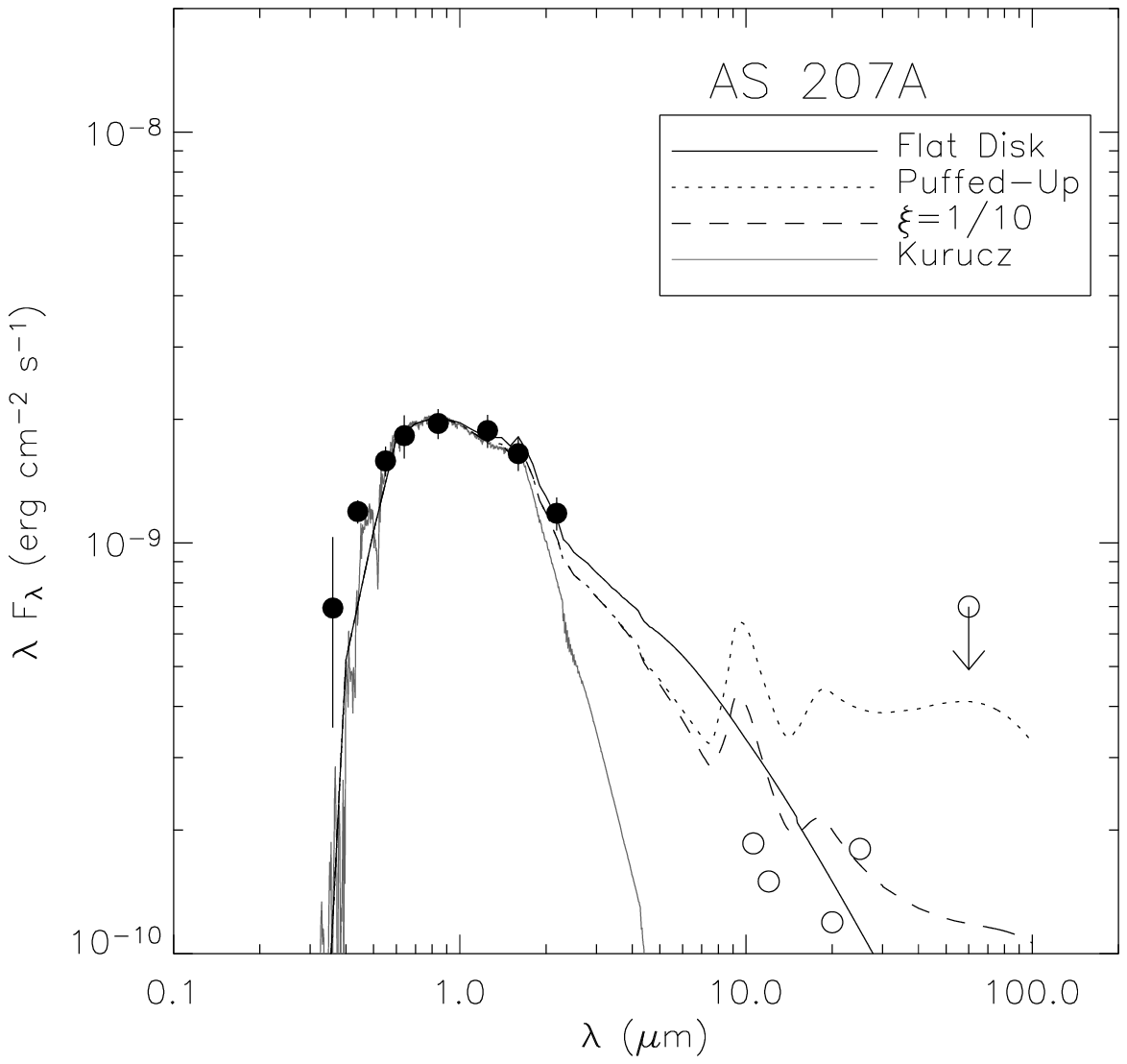}{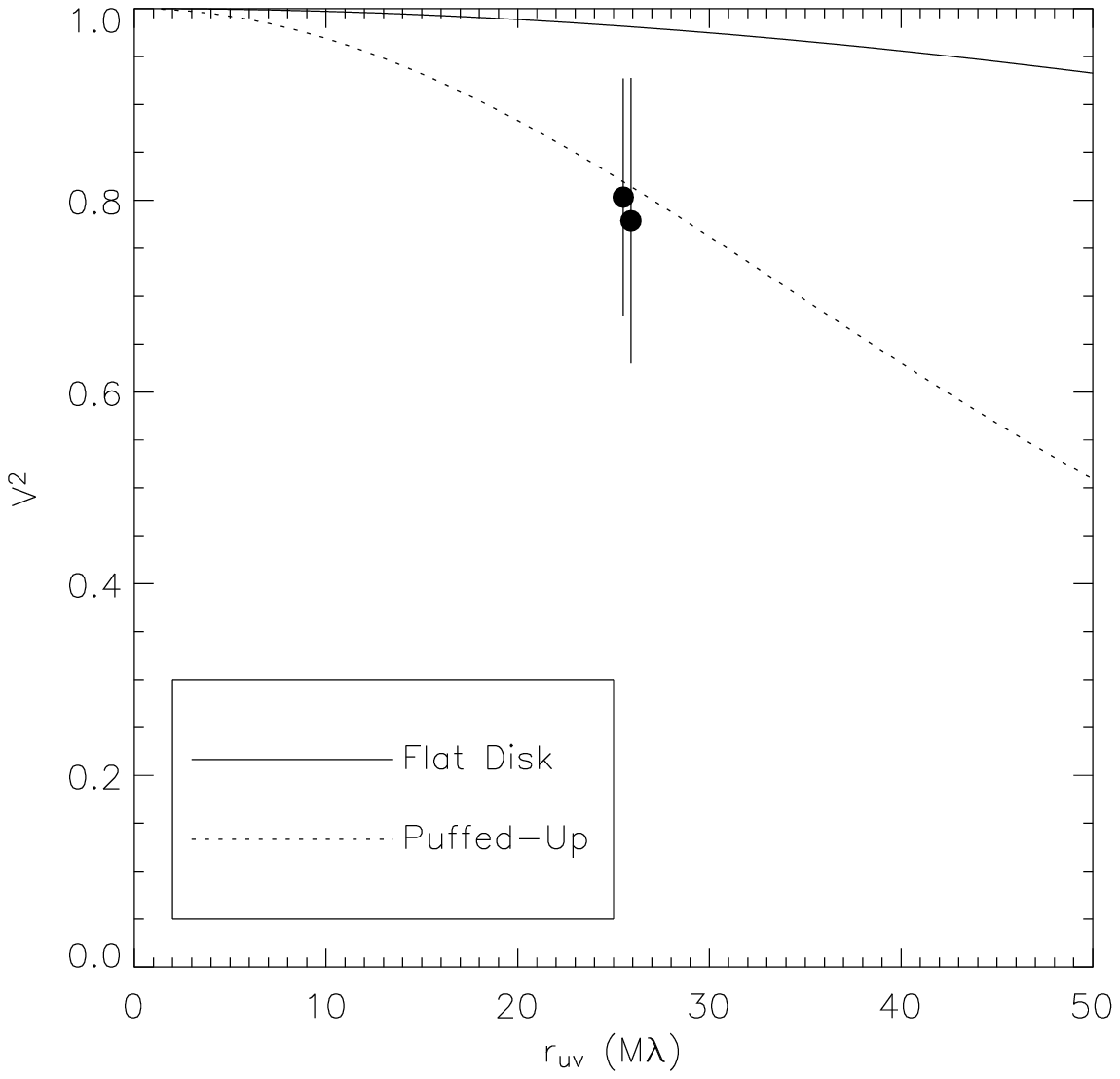}
\caption{(left) Reddening-corrected fluxes for AS 207A 
from optical through near-IR wavelengths
(filled circles), supplemented with longer-wavelength fluxes from the 
literature \citep[open circles;][]{WJ92,JM97,PGS03,KORESKO02}.  
Predicted SEDs for
geometrically flat accretion disks (solid black line), and flared disks with 
puffed-up inner rims (dotted line), as well as the Kurucz model atmosphere
with the stellar parameters determined in \S \ref{sec:stellar}
(solid gray line), are also plotted.  Only $RIJHK$ photometry, probing the
star and inner disk, was used
in the fits. We also plot the SED predicted by a flared disk model
with an intermediate flaring index, $\xi=1/10$ (dashed line).
(right)  Squared visibilities measured with KI, as a function of uv radius,
along with the predictions of different models.  The two curves are labeled
as in the left panel: the solid black line indicates the model 
determined by fitting a flat accretion disk to the $V^2$+SED dataset,
and the dotted line represents the $V^2$ for the best-fit puffed-up inner
disk wall model.  For AS 207A, the puffed-up inner disk model 
provides a superior fit to the data.
The long-wavelength photometry is compatible with an un-flared outer disk.
\label{fig:as207}}
\end{figure}

\begin{figure}
\plottwo{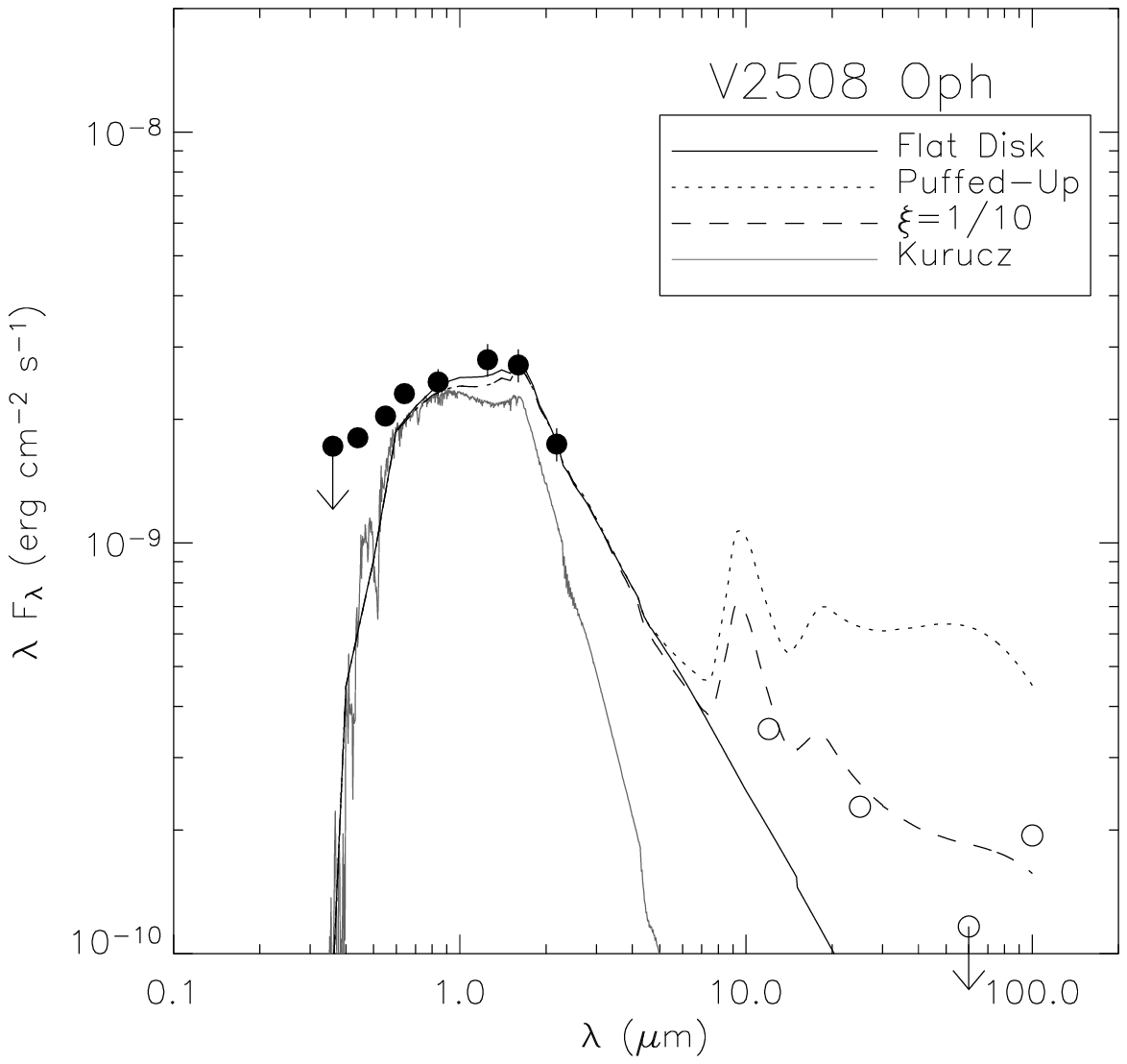}{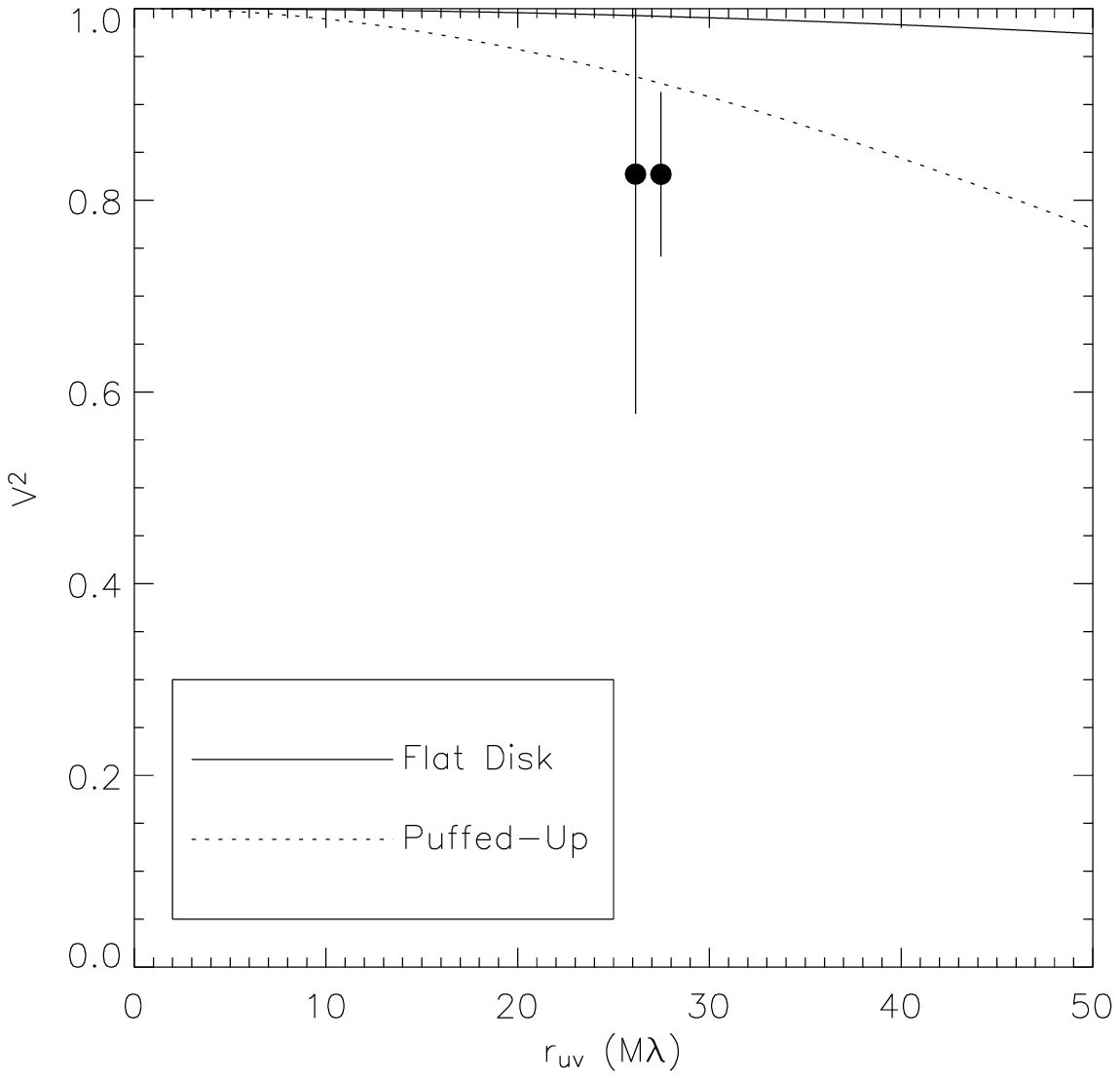}
\caption{SEDs and squared visibilities for V2508 Oph, 
labeled as in Figure \ref{fig:as207}. For this source, the puffed-up inner
disk model provides a superior fit to the SED and $V^2$ data.
The IRAS photometry for this source is more consistent with 
a moderately 
flared disk model (with $\xi \sim 1/10$) than with a flat disk model.
\label{fig:hbc653}}
\end{figure}

\begin{figure}
\plottwo{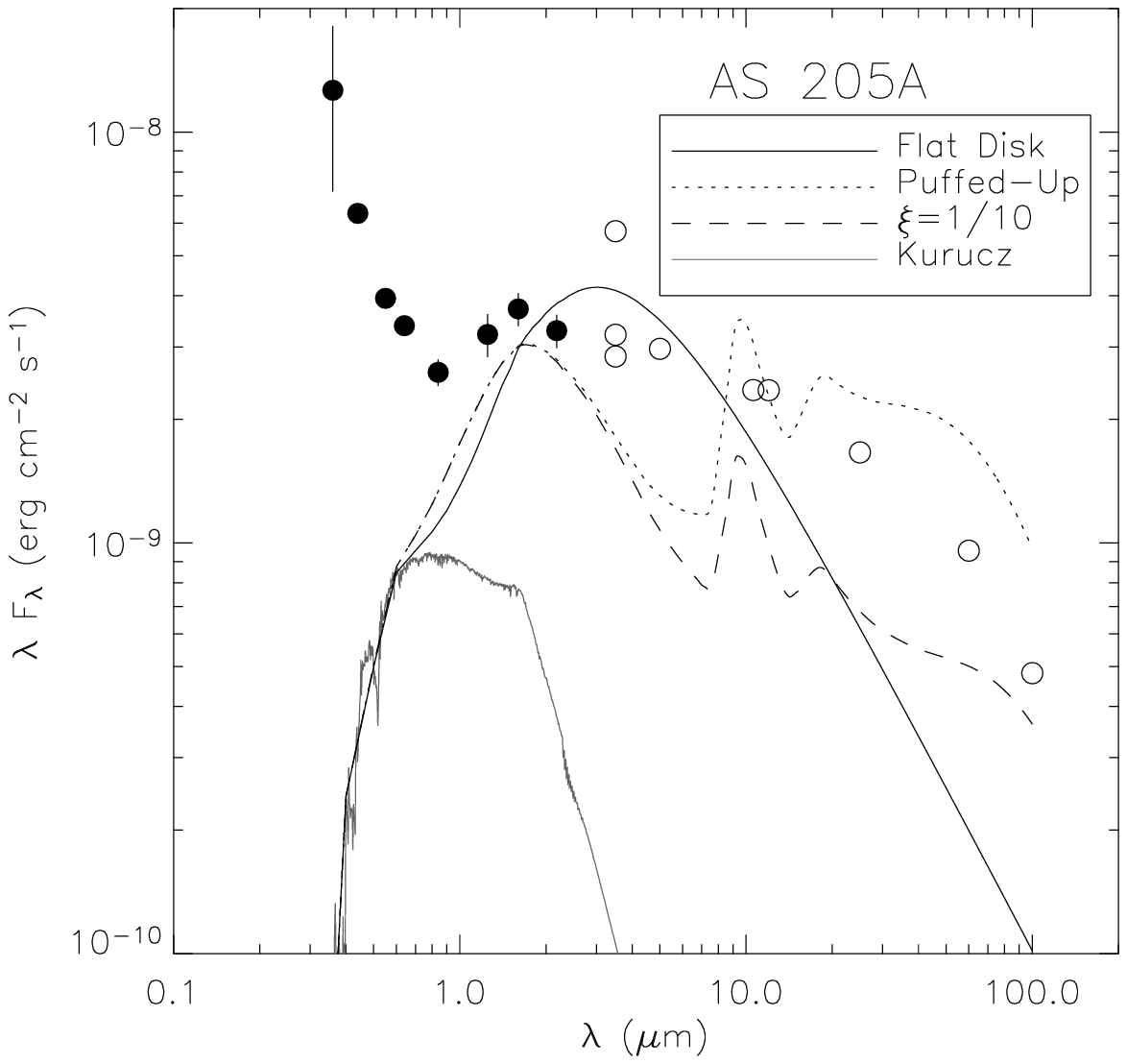}{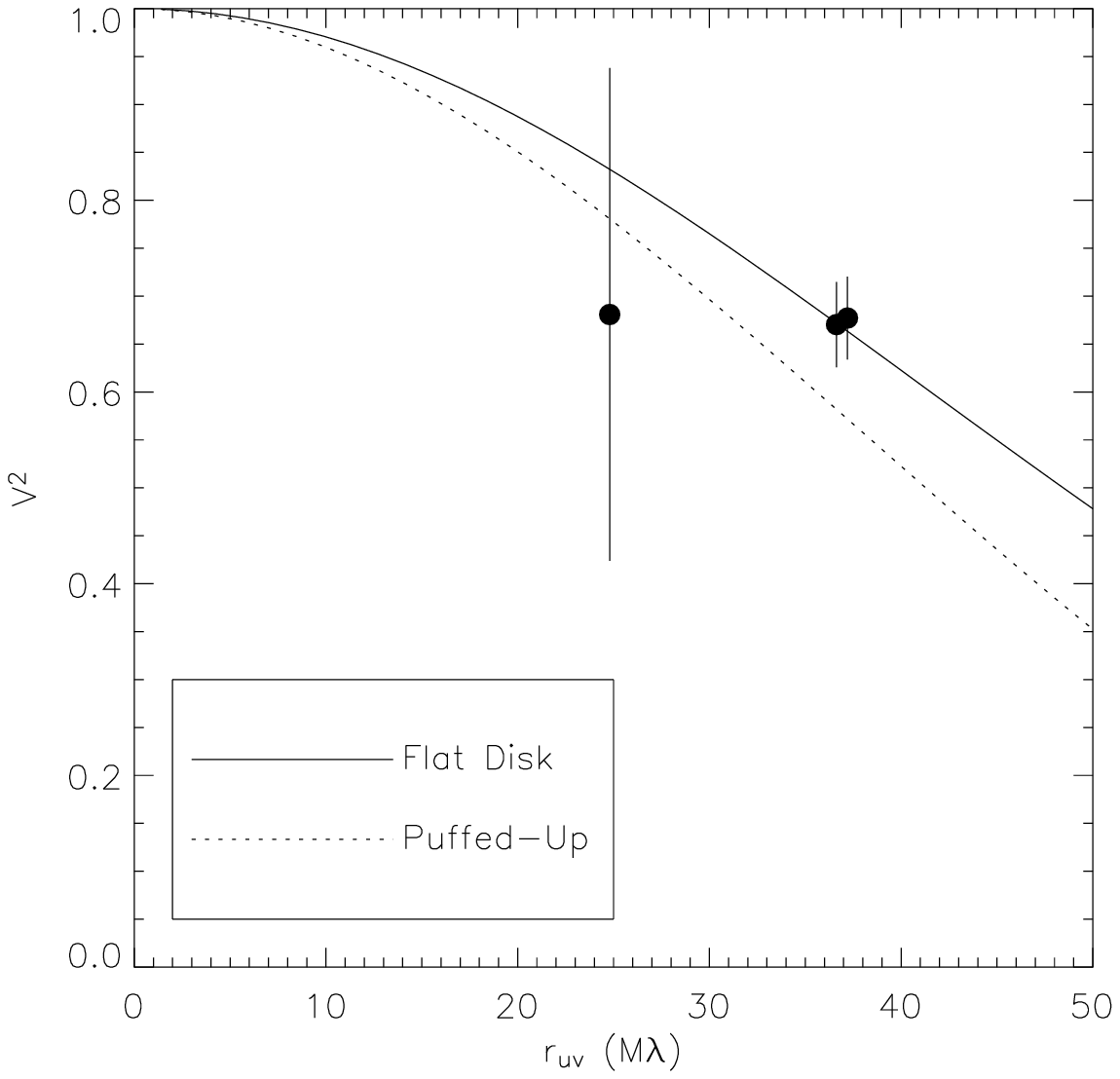}
\caption{SEDs and squared visibilities for AS 205A,
labeled as in Figure \ref{fig:as207}. For this source, the flat disk model
provides a better fit than the puffed up inner disk model to the SED and 
$V^2$ data, although neither model fits very well. These poor fits are
likely due to near-IR emission from hot accretion shocks that is
not accounted for in our models.  The fact that the long-wavelength
photometry lies above the model predictions suggests that flaring and 
accretion heating are important in the outer disk.
\label{fig:as205}}
\end{figure}

\begin{figure}
\plottwo{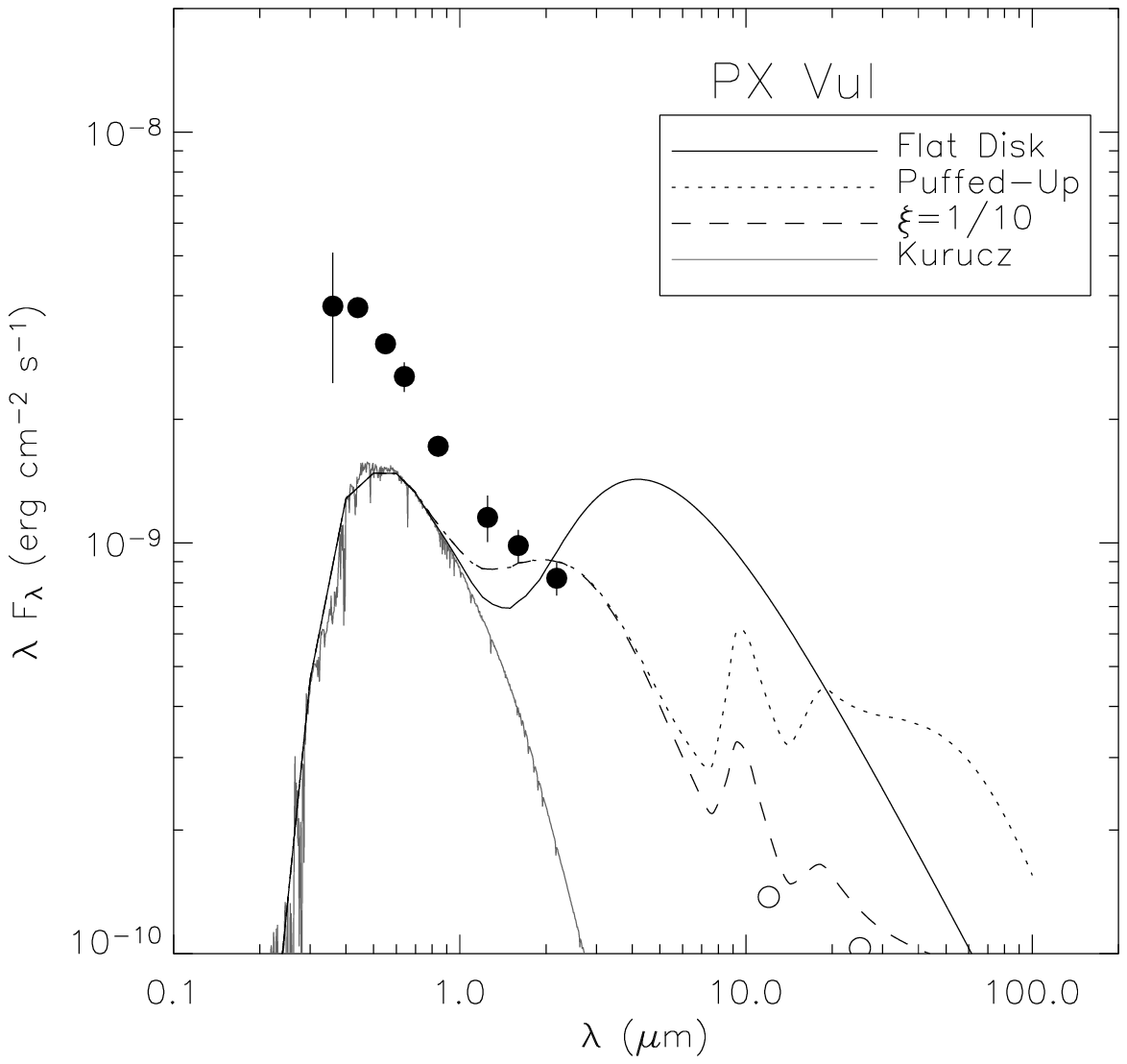}{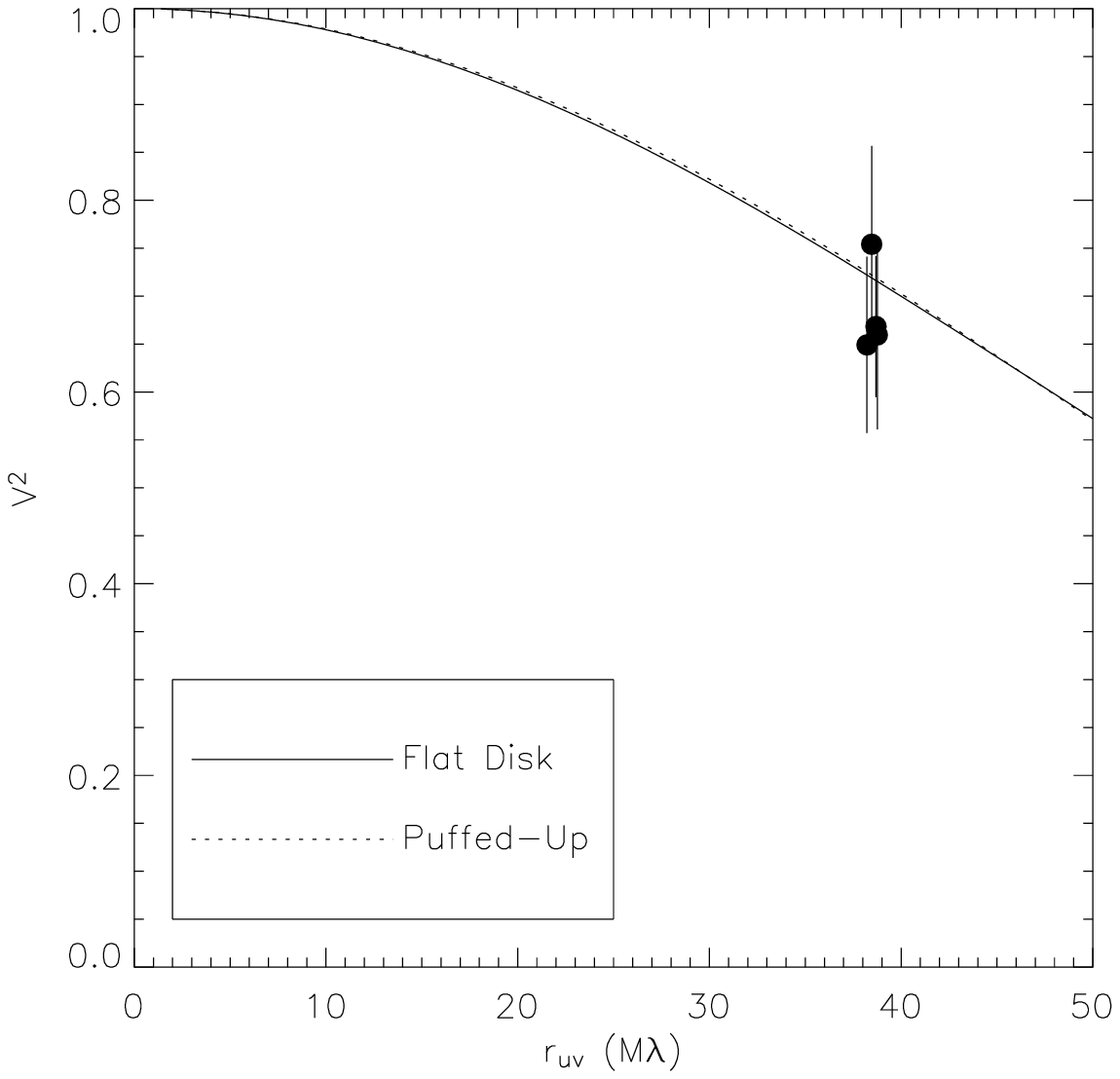}
\caption{SEDs and squared visibilities for PX Vul, 
labeled as in Figure \ref{fig:as207}. For this source, the puffed-up inner
disk model provides a superior fit to the SED and $V^2$ data.
The long-wavelength photometry is compatible with an un-flared outer disk.
Note that for this object, both models fit the $V^2$ data, but lead to
significant deviations in the near-IR SEDs, while for other sources, the SEDs
are similar but the $V^2$ predictions differ 
(Figures \ref{fig:as207}--\ref{fig:as205}); this is due to the
fact that we have more $V^2$ measurements for PX Vul, and the 
combined $V^2$+SED fits are therefore
dominated by the interferometric rather than the photometric data.
\label{fig:hbc293}}
\end{figure}

Calibrated $2.2$ $\mu$m KI visibilities (\S \ref{sec:ki})
and de-reddened SEDs (\S \S \ref{sec:ao}--\ref{sec:optical})
for our sample are shown in 
Figures \ref{fig:as207}--\ref{fig:hbc293}.  $V^2$ values are plotted as a
function of $u-v$ radius, $r_{\rm uv}$, and SEDs use units of
$\lambda F_{\lambda}$.  The SEDs
were constructed from our measured $UBVRIJHK$ photometry, corrected for 
binarity in the case of AS 207 and 
AS 205 (Table \ref{tab:binaries}; \S \ref{sec:stellar}),
and de-reddened using the $A_V$ values in Table \ref{tab:sample}.  
Figures \ref{fig:as207}--\ref{fig:hbc293}
also include longer-wavelength ($> 3$ $\mu$m) photometry from the literature.

%\subsection{Outline}
Measured 2.2 $\mu$m KI visibilities and broadband SEDs
constrain the sizes and temperatures of inner disks around the observed
sources.  However, the near-IR stellar flux contribution to both the SEDs 
and visibilities must be removed before modeling the circumstellar component.
We determine the stellar properties of our sample
in \S \ref{sec:stellar} based on our
spectroscopy and photometry, and use Kurucz models
to determine stellar fluxes at near-IR wavelengths
(see Figures \ref{fig:as207}--\ref{fig:hbc293}; Table \ref{tab:sample2}).
Removing the stellar contributions, we are left with the circumstellar
components of the visibilities and SEDs.

We model these circumstellar components in terms of 1)
a geometrically flat accretion disk and 2) a flared, two-layer,
irradiated disk with a puffed-up inner wall \citep[\S \ref{sec:inner}; as in][]
{EISNER+04}.   For each source, we compute a grid of
models for varying inner disk sizes and temperatures, and find the
``best-fit'' model for which the $\chi^2$ between the model and the data
is minimized.  SEDs and visibilities computed for the best-fit models
are shown in Figures \ref{fig:as207}--\ref{fig:hbc293}, and
best-fit inner disk sizes and temperatures 
are listed in Table \ref{tab:results}.
Longer-wavelength photometry from the literature is
used to qualitatively constrain disk flaring  (\S \ref{sec:flaring}), 
although it is not used in our disk model-fitting.

In order to compare our derived inner disk sizes with those expected from
magnetospheric accretion theory, we use
veiling values and shorter wavelength photometry to constrain mass
accretion rates (\S \ref{sec:mdots}) and thereby determine
magnetospheric truncation radii (\S \ref{sec:rmag}).
Inferred $v \sin i$ values allow estimates of disk co-rotation radii 
(\S \ref{sec:corot}) for comparison purposes.

\subsection{Stellar and Accretion Properties \label{sec:systems}}
\subsubsection{Stellar Properties \label{sec:stellar}}
We determined radial velocities, rotational velocities, spectral types, 
and continuum excesses for our sample from optical spectra 
(Table \ref{tab:sample}), following \citet{WH04}.  Radial velocities and
$v$sin$i$ values are estimated by fitting a parabola to the peak of the
cross-correlation functions, derived using dwarf stars of similar
spectral type.  The spectral types and the optical veiling
levels at $\sim 6500$ \AA $\:$
and $\sim 8400$ \AA\, (defined as $r_{R,I} = F_{\rm excess}
/ F_{\rm photosphere}$) are established simultaneously by comparisons with
artificially veiled dwarf standards\footnote{The lines used
to measure spectral types and veilings are not gravity dependent,
and thus dwarf standards are suitable even though they have higher
surface gravities than our T Tauri sample 
\citep[see][for further discussion]{WH04}.}.  For the binary AS 205, spectral
types are determined for both components from
spatially-resolved spectra.
In this discussion, we focus on AS 205A, but
analysis of AS 205B, itself resolved into a spectroscopic binary, is included 
in Appendix \ref{sec:as205b}.  
The binary AS 207 is spatially unresolved in
the spectral data, preventing extraction of individual components.  However,
the large flux ratio for the AS 207 components (Table \ref{tab:binaries})
suggests that the spectral type of the system is dominated by that
of AS 207A.  AS 207A, V2508 Oph, and AS 205A
have mid-K spectral types, while PX Vul is hotter, with a spectral type of F3.
%Only AS 205A appears to be heavily veiled.

Since the optical
photometry does not resolve the components of AS 205 and AS 207 (\S
\ref{sec:optical}), we measure flux ratios
from adaptive optics or spectroscopic  observations
and use the spectral types for the
components to determine flux ratios at wavelengths where the
pairs are spatially unresolved.  For AS 205, the two Gaussian fits to
echelle spectra (\S 2.4) provide a direct measure of the flux ratios at
$R$ and $I$ bands of 0.16 and 0.24, respectively.  The independently
determined spectral types then provide estimates of flux ratios for the
spatially unresolved $UBV$ measurements.  For AS 207, the two components
are unresolved in the echelle spectra and we employ a less direct
procedure.  We assume that the spatially-resolved J-band measurement (\S
\ref{sec:ao}) probes the photosphere of each component, and then predict
flux ratios at the shorter, spatially unresolved wavelengths using the
measured spectral type for AS 207A (K5) and an assumed
spectral type of M3 for the secondary.  The companion's spectral type is
consistent with both the observed J-band flux ratio and the cooler spectral
types assigned from near-infrared spectroscopy and photometry for the
composite system \citep{DJW03,GM01}.  
%WE MIGHT
%CONSIDER LISTING THE DETERMINED PHOTOMETRY FOR THE COMPONENTS IN TABLE 3,
%PERHAPS UNDER THE UNRESOLVED MEASUREMENT.  IF SO, CONCLUDE THE PARAGRAPH
%WITH A REFERENCE TO THOSE VALUES.

% This maybe an interesting point, but I don't think it goes here:
%
% From Table \ref{tab:binaries}, we also see that the flux ratios between
% components are relatively flat across JHK for AS 205, but steeply
% increasing for AS 207, suggesting a strong IR excess in the secondary of AS
% 205, but little or none in the secondary of AS 207.

Stellar temperatures are assigned based on measured spectral types assuming
a dwarf temperature/spectral type relation \citep[e.g.,][]{HW04}.
Extinctions and stellar luminosities are determined 
by comparing the veiling-corrected $R-I$ fluxes to those expected from Kurucz 
models most similar in temperature, assuming $\log g = 4$ \citep[appropriate
for pre-main sequence stars aged 1-10 Myr; e.g.,][]{PBB95}, the 
extinction relation of \citet{ST91}, and the distances listed
in Table \ref{tab:sample}.  Stellar radii are estimated from the
luminosities and temperatures using the Stefan-Boltzmann equation.
Temperatures, luminosities, radii, and extinctions
 for these stars are listed in Table \ref{tab:sample2}.
We estimate that the assumed stellar temperatures are accurate to
$\pm 100$ K, while luminosities, radii, and extinction estimates
are uncertain by $\sim 20-30\%$.

Stellar masses and ages are estimated by comparing the luminosities
and temperatures with the predictions of pre-main sequence evolutionary
models \citep{SDF00}\footnote{We prefer evolutionary models of \citet{SDF00}
because they span a larger range of stellar masses than those of 
\citet{BARAFFE+98}, and are more consistent with measured dynamical masses 
than \citet{DM97} models.}.
%The observed sample is shown on an H-R diagram in Figure 
%\ref{fig:hr}, along
%with the evolutionary models of \citet{SDF00}.  
These comparisons lead to
masses near 1 M$_\odot$ for the 3 K-type stars, and 2.0 M$_\odot$ for the F3
star.  Stellar ages range from 0.6 to 6.9 Myr (Table \ref{tab:sample2}).  
While considerable
uncertainties in pre-main sequence evolutionary models may lead to large
errors in absolute ages, the relative ages are more secure.
Including adopted uncertainties of 100 K for stellar temperature and 30\% for 
stellar luminosity, AS 207A and V2508 Oph appear to be the youngest stars in 
the sample, while AS 205A is somewhat older, and PX Vul is older still.
We note that the apparent spread in ages may also correspond to different
accretion histories for different sources, which could lead to variations in 
the birthline for pre-main sequence models; thus, the relative ages should be 
treated with some caution.

\subsubsection{Mass Accretion Rates \label{sec:mdots}}
Relative accretion rates for our sample are constrained qualitatively by
H$\alpha$ emission lines in our spectra (Figure \ref{fig:spec}).  Equivalent
widths of H$\alpha$ and full-widths at 10\% of the peak are listed in Table
\ref{tab:sample}.  
These strong, broad profiles suggest on-going accretion in all sources
\citep[e.g.,][]{WB03}, with relatively smaller accretion rates for AS 207A and
V2508 Oph.

Quantitative estimates of mass accretion rates are calculated from
the accretion luminosity generated by infalling material
\citep{GULLBRING+98}:
\begin{equation}
\dot{M} = \frac{L_{\rm acc}R_{\ast}}{G M_{\ast} (1-R_{\ast}/R_{\rm
inner})}.
\label{eq:mdot}
\end{equation}
Here, $L_{\rm acc}$ is the accretion luminosity, $R_{\ast}$ is the stellar
radius, $M_{\ast}$ is the stellar mass, and $R_{\rm inner}$ is the
inner disk radius.  We adopt values of $R_{\rm inner}$ from Table
\ref{tab:results} (using puffed-up inner disk sizes determined from
combined $V^2$+SED analysis; see \S \ref{sec:inner});  these are
inner radii of the dust disks, and may be somewhat larger than the inner
gas radii relevant for this formula (as discussed in \S \ref{sec:drmag}), 
which would consequently lead to
larger inferred mass accretion rates.  

The accretion luminosity, $L_{\rm acc}$, is estimated by applying
a bolometric correction factor to a flux excess
measured over a limited wavelength range.
We calculate accretion luminosities using two methods, one based
on measured veiling at $R$-band \citep[e.g.,][]{HK03,WH04}, and the other 
based on measured $U$-band excess emission \citep{GULLBRING+98}.  
For the first method, $R$-band excess luminosities are calculated
from the measured veilings and then converted into accretion luminosities
using a bolometric correction of 35.  The bolometric correction factor
is highly uncertain, and probably introduces uncertainties of a factor
of $\sim 3$ in the computed accretion luminosities.  We also calculate the
accretion luminosity from the observed $U$-band excess luminosity ($L_U$)
following \citet{GULLBRING+98}:
\begin{equation}
\log ({L_{\rm acc}}/{L_\odot}) = 1.09^{+0.04}_{-0.18} 
\log ({L_U}/{L_\odot}) + 0.98^{+0.02}_{-0.07}.
\label{eq:lacc}
\end{equation}
Although the accretion luminosities calculated using Equation \ref{eq:lacc}
use a smaller bolometric correction than for the first method \citep[due
to the assumed high temperature of the accretion excess;][]{CG98}, the
large photometric uncertainties for our $U$-band data (Table \ref{tab:phot})
introduce errors of a factor of $\sim 2$.  We find that accretion estimates
based on $U$-band fluxes are typically higher than those computed
from $R$-band measurements, although the two estimates are consistent to
within a factor of 2.  Since the accretion luminosities
estimated from both methods have large error bars, we adopt the mean of
the two values in our analysis.  
Inferred accretion luminosities for our sample range from 0.4 $L_\odot$
to 25.0 $L_\odot$, and are listed in Table \ref{tab:sample2}.  

Using these adopted values for $L_{\rm acc}$, we calculate mass accretion 
rates from Equation \ref{eq:mdot}.  For our sample, $\dot{M}$ is between
$3.2 \times 10^{-8}$ M$_\odot$ yr$^{-1}$ and 
$1.3 \times 10^{-6}$ M$_\odot$ yr$^{-1}$  (Table \ref{tab:sample2}).
The large uncertainties for $L_{\rm acc}$ lead to
accretion rates that are probably uncertain by a factor of 2-3.

\subsubsection{Magnetospheric Radii \label{sec:rmag}}
The expected radius of magnetospheric truncation, $R_{\rm mag}$, is 
determined by the balance of forces between infalling (accreting)
material and the stellar dipole field \citep{KONIGL91}:
\begin{equation}
\frac{R_{\rm mag}}{R_{\ast}} = 2.27 \left[\frac{(B_0/{\rm 1 \: kG})^4 
(R_{\ast}/{\rm R_{\odot}})^5}{(M_{\ast}/{\rm M_{\odot}}) 
(\dot{M}/10^{-7} {\rm M_{\odot} \: yr}^{-1})^2}\right]^{1/7}.
\label{eq:rmag}
\end{equation}
With the stellar parameters determined in \S \ref{sec:stellar},
the accretion rates calculated in \S \ref{sec:mdots}, and assuming
a typical magnetic field strength for T Tauri stars of 2 kG
\citep{JVG03}, we calculate $R_{\rm mag}$ for our sample.
Our values for $R_{\rm mag}$ range from 0.03 to 0.13 AU
(Table \ref{tab:sample2}).  Propagating the assumed uncertainties 
for $R_{\ast}$, $M_{\ast}$, and $\dot{M}$, and adopting an uncertainty of 30\%
for $B_0$, we estimate that the
magnetospheric radii are uncertain by $\sim 30\%$.

\subsubsection{Co-Rotation Radii \label{sec:corot}}
The co-rotation radius is the radius at which the Keplerian
orbital period in the disk equals the stellar rotation period.
We derive co-rotation
radii for stars with $v \sin i$ measurements (Table \ref{tab:sample})
according to:
%\begin{equation}
%R_{\rm co-rotation} = G M_{\ast} \left(\frac{\sin i}{v \sin i}\right)^2
%\le G M_{\ast} \left(\frac{1}{v \sin i}\right)^2.
%\label{eq:rco}
%\end{equation}
\begin{equation}
R_{\rm co-rotation} = (G M_{\ast})^{1/3} \left(\frac{R_{\ast}}{v}
\right)^{2/3} \le (G M_{\ast})^{1/3} \left(\frac{R_{\ast}}{v \sin i}
\right)^{2/3} .
\label{eq:rco}
\end{equation}
Here, $M_{\ast}$ is the stellar mass, $R_{\ast}$ is the stellar radius,
$v \sin i$ is the projected rotational velocity of the star, and $i$ is
the inclination of the system.
For AS 207A, where there is a reported photometric period, $\tau=6.53$ days
\citep{SH98,BOUVIER90}, the inclination (and hence the
co-rotation radius) can be determined explicitly.
%\begin{equation}
%i = \sin^{-1}\left(\frac{\tau \: v \sin i}{2 \pi R_{\ast}}\right) =
%43^{\circ}.
%\end{equation}
For the remaining sources, without known rotation periods \citep{SH98}, 
we derive upper limits.  Co-rotation radii and upper limits range from
0.03 to 0.09 AU, and are listed in
Table \ref{tab:sample2}.  Propagating the uncertainties in $M_{\ast}$,
$R_{\ast}$, and $v \sin i$ (assuming stellar mass and radius are uncertain
by $\sim 30\%$), we estimate that the derived co-rotation radii are
uncertain by approximately 20\%.

\subsection{Near-IR Stellar and Excess Fluxes \label{sec:components}}
The measured 2.2 $\mu$m visibilities and near-IR SEDs contain information 
about inner circumstellar disks as well as the central stars, and 
distinguishing the stellar and excess fluxes is crucial to accurate modeling of
the disks. Because the stellar SED peaks closer to 2.2 $\mu$m for our current 
sample than for the hotter stars analyzed in \citet{EISNER+04},
this step is especially critical here.
In this Section, we discuss our procedure for removing the stellar component 
from the $V^2$ and SED data, and in \S \ref{sec:inner}, we
model the circumstellar component and determine disk parameters.

The measured SED at near-IR wavelengths is simply the sum of the stellar and
disk fluxes.  Our 2.2 $\mu$m visibilities are given by
\begin{equation}
V^2_{\rm meas} = \left(\frac{F_{\ast}V_{\ast} + F_{\rm D}V_{\rm D}}
{F_{\ast} + F_{\rm D}}\right)^2 \approx
\left(\frac{F_{\ast} + F_{\rm D}V_{\rm D}}
{F_{\ast} + F_{\rm D}}\right)^2,
\label{eq:decomp}
\end{equation}
where $F_{\ast}$ is the stellar flux, $V_{\ast} \approx 1$ are the visibilities
due to the unresolved central star, $F_{\rm D}$ is the circumstellar disk 
flux, and $V_{\rm D}$ are the visibilities due to the disk.
A knowledge of the stellar flux at near-IR wavelengths is critical for modeling
the circumstellar components of both the 2.2 $\mu$m visibilities and the
near-IR SEDs.  This flux can be estimated using the
stellar parameters from \S \ref{sec:stellar} and
extrapolating the veiling-corrected flux measured at
optical wavelengths using a Kurucz model.
The excess 2.2 $\mu$m flux due to the compact circumstellar disk
is the difference between the de-reddened observed flux 
and the Kurucz model.  These 2.2 $\mu$m
stellar and excess fluxes are listed in Table \ref{tab:sample2}.

The inferred stellar and excess fluxes are somewhat 
uncertain, leading to uncertainties in the derived disk parameters.  
For AS 205A and PX Vul, where the excess flux dominates the emission,  
uncertainties in the relative fluxes will have a small effect on 
disk parameters.  However, when the stellar and excess fluxes
are comparable, as for AS 207A and V2508 Oph, there can be
significant errors in the fitted disk
sizes.  For example, 30\% errors in 2.2 $\mu$m stellar flux lead to size 
errors of 25\% and 23\% for AS 207A and V2508 Oph, but only 2\% and 7\% for
AS 205A and PX Vul. 

In addition to the central star and circumstellar disk, there may be other
contributions to the visibilities and SEDs.  Emission on scales
between the $\sim 5$ mas KI fringe spacing and the $\sim50$ mas field of view,
for example, due to thermal or scattered emission from residual envelopes, 
will be resolved out and will lower the overall measured
visibilities and lead to larger inferred sizes.  If
the extended emission has a similar spectrum to the star, as expected for
scattering from large dust grains, it will not affect the SED.  
Extended emission is difficult to constrain, 
since observations that can resolve faint emission on angular scales 
smaller than  50 mas are virtually non-existent. 
However, previous analysis of the visibilities and SEDs of more luminous
Herbig Ae/Be stars indicates small ($\la 1\%$) contributions from extended 
emission on $\la 1''$ scales \citep{EISNER+04}.  For KI, which has a field 
of view of only 50 mas, we expect the contribution from extended emission 
for our less luminous T Tauri star sample to be even smaller.  

Near-IR emission may also arise from hot gas in accretion shocks at the
stellar photosphere
\citep[e.g.,][]{CG98,GULLBRING+00}, or hot optically thin gas in the
inner disk \citep[interior to the dust truncation radius; e.g.,][]{AKESON+05}.
Since these hot gas components are compact compared to emission from the
circumstellar dust disk, they would tend to raise the visibilities
compared to those predicted by our Equation \ref{eq:decomp}; 
i.e., lead to smaller inferred
disk sizes.  In contrast, for the measured SEDs hot gas would contribute 
extra emission, leading to {\it larger} inferred disk sizes.
%These additional components would modify Equation \ref{eq:decomp} to
%\begin{equation}
%V^2_{\rm meas} = 
%\left(\frac{F_{\ast} + F_{\rm D}V_{\rm D} + F_{\rm gas}V_{\rm gas}}
%{F_{\ast} + F_{\rm D} + F_{\rm ext} + F_{\rm gas}}\right)^2,
%\label{eq:starcs}
%\end{equation}
%where $F_{\rm ext}$ is the 
%flux in the extended component ($V_{\rm ext} \approx 0$ are the 
%visibilities due to this over-resolved component) and $F_{\rm gas},
%V_{\rm gas}$ are the flux and visibilities due to the hot gas component.
%Similarly, no emission from hot gas was required to explain the
%Herbig Ae/Be star observations, suggesting that this component is
%probably also relatively unimportant for modeling the near-IR data.
We expect that the fraction of near-IR emission
from an 8000 K accretion shock \citep{CG98} will be relatively small
compared to peak emission from a 1000-2000 K disk.
However, emission from hot gas may
cause a  measurable effect on the 2.2 $\mu$m visibilities and near-IR SEDs for 
sources with extremely high accretion rates (see \S \ref{sec:as205}). 
%Furthermore, this hot emission
%will have a large effect on short-wavelength photometry, and we therefore
%exclude $UBV$ photometry when modeling the inner disk.
%Since we cannot constrain these additional emission components, 
%we assume that at 2.2 $\mu$ m, $F_{\rm gas}=0$.  Thus, we use Equation
%\ref{eq:decomp} in the analysis below.

In the analysis below, we assume that near-IR emission from extended dust
or hot gas is insignificant compared to the stellar and disk emission.
Thus, we model the SED using a Kurucz stellar atmosphere plus thermal
emission from a disk,
and use Equation \ref{eq:decomp} to model the measured visibilities.

\subsection{Modeling Inner Disk Structure \label{sec:inner}}
Equipped with the stellar and circumstellar contributions to the
visibilities and SEDs (\S \ref{sec:components}), we fit the circumstellar
components using the two simple disk models described in detail by 
\citet{EISNER+03,EISNER+04}: 1)  a geometrically flat 
accretion disk with a temperature profile $T(R) \propto R^{-3/4}$ \citep{LP74},
truncated at an inner radius, $R_{\rm in}$; and 2) a flared, 
irradiated, two-layer disk \citep{CG97} incorporating a puffed-up inner 
disk wall at $R_{\rm in}$ \citep{DDN01}.  The main difference between
the two models is the angle of incidence of stellar radiation:
for the flat disk, stellar radiation is incident at glancing angles,
while the puffed-up inner disk and flared outer surface intercept
starlight at more normal angles, leading to additional disk heating.
For the second model, the near-IR emission is dominated by the puffed-up 
inner rim, and the emission appears essentially ring-like at the 
2.2 $\mu$m wavelength of our interferometric observations.
While the geometrically-thin
disk model assumes blackbody emission, we assume the opacity for the
puffed-up inner disk model is due to astronomical silicate dust \citep{DL84}.

For each model, 
the parameters relevant for the inner disk structure are the inner radius, 
$R_{\rm in}$, and the
temperature of the disk at the inner radius, $T_{\rm in}$.
Temperatures at other disk radii are specified by these parameters and the
assumed temperature profiles for the disk models.
%Stellar flux is included in these models using a Kurucz model atmosphere
%with values of  $T_{\ast}$ and $R_{\ast}$ determined in 
%\S \ref{sec:stellar}.  
The $V^2$ and SED data (and associated error bars) 
used in the modeling are shown in Figures 
\ref{fig:as207}-\ref{fig:hbc293}.  
We use only $RIJHK$ photometry, and de-veil the $RI$ fluxes, so that
the data traces only the inner disk, uncontaminated by 
hot accretion excess emission which can dominate at shorter wavelengths.

With the limited amount of data available (2-4 visibilities, and 5
photometric points for each source), we consider only face-on disk
models here.  As discussed in \citet{EISNER+04}, including inclination
in the models may affect the results. Unless the baseline position
angle corresponds with the major axis of an inclined disk, the
size inferred from $V^2$ measurements for a face-on model would be 
underestimated with respect to the inferred size for an inclined model.  
Similarly, the face-on assumption would lead to an underestimate of disk
size from SED data, 
since a face-on disk produces more near-IR flux than an inclined disk of the 
same size.   Further interferometric observations, probing a 
range of position angles, are necessary to constrain the parameters 
of inclined disk models.  However, assuming our sources are not close to
edge-on (reasonable given that the central stars are un-obscured),
inclination effects will not substantially alter the results presented
here for face-on disk models.

We fit our 2.2 $\mu$m KI visibilities and SED data simultaneously and determine
the best-fit parameter values, $R_{\rm in}$ and $T_{\rm in}$,
by calculating $\chi^2$ for models over a grid of
inner radii and temperatures.  In practice, we fit for the directly observable
angular (rather than linear) size of the inner radius, $\theta_{\rm in}$;
best-fit values are converted to linear sizes using the distances in Table
\ref{tab:sample}.  We consider $\theta_{\rm in}$ 
ranging from 0.1 to 10 mas in increments of 0.01 mas (spanning the
approximate angular resolution of KI), and 
$T_{\rm in}$ ranging from 1000 to 2500 K in 100 K
increments \citep[bracketing values expected for dust sublimation; e.g.,][]
{SALPETER77,POLLACK+94}.  For each model, we calculate the $\chi^2$ of the
combined $V^2$+SED dataset, where each data point is weighted by its
measurement uncertainty, and we find the
inner disk size and temperature for which $\chi^2$ is minimized.  
1-$\sigma$ uncertainties on the best-fit parameters are determined
in the standard way \citep[e.g.,][]{EISNER+04}. Best-fit parameters,
1-$\sigma$ uncertainties, and reduced $\chi^2$ values are listed in Table 
\ref{tab:results}.  
Puffed-up inner disk models generally provide small $\chi^2$ values
(indicating good fits to the data), with
inner disk sizes and temperatures ranging from 0.12 to 0.32 AU 
and 1000 to 1800 K,
respectively.  In contrast, flat disk models fit the data poorly, and
suggest smaller inner disk sizes (0.02-0.22 AU) and higher temperatures
(1400-2400 K). 

The poor fits of flat disk models to the data are consistent with
previous observations which showed that sizes determined from 
near-IR interferometry are often larger than those determined from SED
modeling \citep[e.g.,][]{AKESON+00,MST01,EISNER+03}.  
We investigate this issue here by 
fitting the visibility and SED data separately.
For the SEDs, we use the same
procedure and parameter grid as for the combined analysis.  For the 
visibilities, where we have fewer data points, 
we fit only for $\theta_{\rm in}$,
assuming the value of $T_{\rm in}$ determined from the combined analysis.
The best-fit parameter values, uncertainties, and reduced $\chi^2$ values
are included in Table 
\ref{tab:results}.  We note that for the small numbers of visibility
measurements, the reduced $\chi^2$ are often very small, indicating poor
constraints on the models.  The results indicate that the $V^2$ or SED data
individually can be fit reasonably well with either model, although
the puffed-up inner disk model provides somewhat lower $\chi^2$ values
for the SED fits.  For most sources, the inner size for flat
disk models inferred from the visibilities is $\ge 5$ times larger
than inferred from the SEDs.  The puffed-up inner disk models,
in contrast, find best-fit sizes from the $V^2$ or SEDs
generally consistent within the 1$\sigma$ uncertainties 
(Table \ref{tab:results}).

From the SEDs alone, we cannot distinguish between flat and puffed-up
inner disk models, but combined $V^2$+SED
analysis shows that puffed-up inner disk models are preferred
 (Table \ref{tab:results}).
Qualitatively, this additional constraint comes from the spatially
resolved information contained in the $V^2$ data.
While near-IR SEDs constrain both the temperature and size
of the inner disk, these parameters are degenerate with the 
spatial and temperature profiles, and thus SED fits are not unique;
one can find a suitable fit for either the geometrically thin
or puffed-up inner disk models by varying $T_{\rm in}$ and $R_{\rm in}$.
Combining SEDs with even a limited amount of interferometric data,
we can measure {\it directly} the size of the inner disk, thereby
breaking the degeneracy inherent in SED-only modeling and enabling us to
distinguish between puffed-up and geometrically flat inner disk models.
  
The measured sizes discussed above are determined directly from the data.
Since we separated the circumstellar components of the visibilities and
SEDs from the stellar contributions in \S \ref{sec:components}, the
measured disk sizes  do not depend on stellar
properties or disk accretion rates; i.e.,  $R_{\rm in}$  and $T_{\rm in}$
are chosen simply to provide the best fit to the observations.
Thus, the inner disk structure for our best-fit models is fully
specified by $R_{\rm in}$, $T_{\rm in}$, and the assumed temperature profiles.
Below, we investigate whether the stellar and accretion luminosities
in these sources can provide sufficient disk heating to match the
measured inner radii and temperatures, providing an additional test of whether 
the measured sizes are consistent with the physical models.

For flat disk models, the expected temperature in the disk at 1 AU
depends on heating by both stellar radiation and viscous dissipation
of accretion energy \citep{LP74},
\begin{equation}
T_{\rm 1 AU} = \left[2.52 \times 10^{-8} \left(\frac{R_{\ast}}{\rm R_{\odot}}
\right)^3 T_{\ast}^4 + 5.27 \times 10^{10} \left(\frac{M_{\ast}}{\rm M_{\odot}}
\right) \left(\frac{\dot{M}}{10^{-5} \: {\rm M_{\odot}} \: {\rm yr}^{-1}}
\right)\right]^{1/4} .
\label{eq:racc2}
\end{equation}
Thus, the expected disk temperature depends on 
$T_{\ast}$, $R_{\ast}$, $M_{\ast}$,
and $\dot{M}$.  Using the value of $T_{\rm 1 AU}$ computed for
our inferred stellar and accretion parameters (\S \ref{sec:systems}),
we predict the radius in the disk where $T=T_{\rm in}$ (where
$T_{\rm in}$ is determined from combined $V^2$+SED analysis),
\begin{equation}
\frac{R_{\rm flat}}{\rm AU} = 
\left(\frac{T_{\rm 1 AU}}{T_{\rm in}}\right)^{4/3}.
\label{eq:racc1}
\end{equation}
For the passive disk model with a puffed-up inner wall,
the expected radius where $T=T_{\rm in}$ depends on the total luminosity
incident on the inner disk,
\begin{equation}
R_{\rm puffed-up} = 
\sqrt{\frac{L_{\ast} + L_{\rm acc}}{4 \pi \sigma T_{\rm in}^4} (1+f)}.
\label{eq:rpuff}
\end{equation}
Here, $f$ is the ratio of the inner disk width to the inner radius,
which we have assumed to be 0.10 \citep{DDN01}.  Equation
\ref{eq:rpuff} includes the effects of accretion luminosity, $L_{\rm acc}$,
in addition to the stellar luminosity \citep{MUZEROLLE+03}.

Expected inner radii for the two models,
with and without accretion heating, are listed in Table 
\ref{tab:predictions}.  We compare these predictions with our measured sizes:
for the puffed-up inner disk model, we use sizes measured from
$V^2+$SED data, while for the flat disk model, which provides poor
fits to combined datasets, we use the inner disk sizes measured from
$V^2$-only data.
Measured inner
disk sizes are roughly consistent with expectations for puffed-up inner 
disk models based on the stellar parameters determined in \S \ref{sec:stellar}.
Moreover, for the high-accretion rate source AS 205A, the predicted
size is more consistent with the measured size when accretion 
luminosity is included, demonstrating the importance
of accretion in the disk structure for this object.
In contrast, for AS 207A, V2508 Oph, and PX Vul, the predicted sizes
of puffed-up inner disk models with $\dot{M}=0$ are compatible with the
measured sizes (Table \ref{tab:predictions}), suggesting that stellar
irradiation is the dominant effect in puffing up the inner disk edges.
Expected sizes for the flat disk model are smaller than  measured values 
for all sources except AS 205A.   These results are compatible with the fact 
that the puffed-up inner disk models generally fit the visibility and SED 
data better than the flat disk models.

%Since these estimates depend on $T_{\rm ast}$, $L_{\rm ast}$, and
%$\dot{M}$, all of which are somewhat uncertain, the predicted sizes
%will also have some associated errors bars. We adopt an uncertainty in the 
%predicted sizes of 20\%.

The best-fit inner disk sizes (Table \ref{tab:results}) 
are larger than both the magnetospheric and co-rotation radii calculated
in \S \ref{sec:systems} (Table  \ref{tab:sample2}).  We illustrate
this graphically in Figure \ref{fig:skyplots}, where we plot the 2.2 $\mu$m
brightness distributions for our best-fit puffed-up inner disk models,
and indicate the positions of magnetospheric and co-rotation radii with
dotted and dashed lines, respectively. The discrepancy
between measured sizes and magnetospheric/co-rotation radii 
is relatively small for some sources ($\la 2$), and large for others ($>5$).
The magnitude of these discrepancies depends to some extent on our
assumptions.  However, more realistic models including inclined disks
and potentially lower stellar magnetic fields would actually exacerbate
the differences between measured sizes and magnetospheric radii \citep[magnetic
fields substantially higher than the assumed 2 kG are unlikely;][]{JVG03}.

\subsection{Large-Scale Disk Structure \label{sec:flaring}}
In \S \ref{sec:inner}, we modeled our $RIJHK$ photometry and 2.2 $\mu$m
visibilities, and determined inner disk radii and temperatures
for our sample.  These values of $R_{\rm in}$, $T_{\rm in}$ 
also provide the normalization of the temperature profiles (for our two
simple models) in the outer disk regions.
Here, we combine our measurements and modeling of the inner 
disk with longer-wavelength photometry 
\citep[3-100 $\mu$m;][]{WJ92,JM97,PGS03,KORESKO02}
which probes larger disk radii.
Due to source variability and multiple
sources within the large IRAS beam (as seen in 2MASS images),
the uncertainties in this long-wavelength photometry
are likely $\ga 10\%$.  
Despite  these uncertainties, the long-wavelength fluxes still yield rough
constraints on outer disk structure.  

We quantify disk flaring by how the height of the
disk increases with radius: $H/R \propto R^{\xi}$.
For a flat disk, $\xi=-1$, while for a fully flared disk in hydrostatic 
equilibrium, $\xi=2/7$
\citep{CG97}. These two extremes correspond to the flat and puffed-up
disk models used above.  However, dust settling and/or grain growth could
result in other values for $\xi$ \citep{DD04b,DD04}.
Here, we also consider the case of a
somewhat, but not fully, flared disk with $\xi=1/10$.  
Comparison of the un-flared, somewhat flared, 
and fully flared models with the data 
give a qualitative measure of the degree of flaring.
Inclination effects, which are not included in our 
face-on models, will also suppress the long-wavelength flux (due to the
smaller projected surface area), mimicking the effects of flatter disks.
Thus, we do not attempt
to determine $\xi$ exactly, instead maintaining a qualitative approach.

The outer disk geometry, as illustrated
by the long-wavelength photometry, seems to vary from source to source
(Figures \ref{fig:as207}--\ref{fig:hbc293}).
In some sources (AS 207A, PX Vul), flatter outer disks are consistent with the 
data, while other objects (V2508 Oph, AS 205A) require significant outer disk 
flaring to explain the data.

\section{Results for Individual Sources \label{sec:results}}
Figures \ref{fig:as207}--\ref{fig:hbc293} show the
flat (solid lines) and puffed-up (dotted lines) disk
models that provide the best fits to combined $V^2$+SED datasets.
For each source see Table \ref{tab:results} for best-fit 
$R_{\rm in}$, $T_{\rm in}$, and reduced $\chi^2$ values of these models.
In general, puffed-up inner disk models with inner temperatures
ranging from 1000-1800 K provide good fits to the data, while 
flat disk models provide poor fits to the SED and $V^2$
data.

\subsection{AS 207A \label{sec:as207}}
AS 207A shows a weak near-IR excess and a mass accretion rate lower than
other sources in our sample.  The visibility and SED data are
more consistent with the predictions of a cooler,
puffed-up inner disk model than with a flat disk model
($\chi_r^2 = 0.21$ versus $1.07$;  Figure \ref{fig:as207}).  
Best-fit inner disk sizes and temperatures are
$\sim 0.25$ AU and 1000 K for the puffed-up inner disk model.   
This size is approximately two times larger than the
calculated magnetospheric and co-rotation radii
(Figure \ref{fig:skyplots}).

The IRAS photometry for AS 207A is compatible with flat outer disk models.
Although AS 207A has a binary companion, 
the steeply increasing flux ratio across
$J$,$H$, and $K$ (Table \ref{tab:binaries}) suggests that the companion
contributes little to the long-wavelength flux.

\subsection{V2508 Oph \label{sec:v2508oph}}
For V2508 Oph, a source with 
a relatively small accretion rate,
the puffed-up disk model provides a
better fit to our $V^2$ and SED data than the
flat disk model ($\chi_r^2=0.98$, compared with 1.41).  
While the puffed-up inner disk size determined from the 
visibilities (0.20 AU) is somewhat larger than that determined from the 
SED (0.07 AU), the discrepancy has $<1\sigma$ significance, and the
fit to the combined $V^2$+SED dataset produces a reasonable $\chi_r^2$
value (Figure \ref{fig:hbc653}; Table \ref{tab:results}).  
Furthermore, the measured inner disk size agrees well with 
(but is slightly larger than) the magnetospheric
truncation radius calculated in \S \ref{sec:rmag} and the upper limit on 
co-rotation radius determined in \S \ref{sec:corot} 
(Figure \ref{fig:skyplots}).
The long-wavelength photometry is compatible with
an outer disk that is somewhat flared ($\xi \sim 1/10$).

%The size determined from the visibilities is also somewhat larger than 
%predicted by simple models (Table \ref{tab:predictions}, although
%the discrepancy is again only significant to $<1\sigma$).  

%The disagreement between the SED and visibility models may be due to
%several factors.  One possibility is an error in stellar parameters:
%as mentioned above, an error of 30\% in the estimated stellar flux can
%lead to an error of 23\% in the measured size of V2508 Oph.  Another
%possibility is that more complex models are necessary.
%For example, highly-inclined disk models would decrease the discrepancy between
%the SED and visibility models.  Alternatively, there may be some emission at
%large radii, for example from a remnant envelope or a companion within 50 
%mas, that is causing somewhat lower visibilities than expected.  These
%more complex models are not warranted by the data presented here, but should
%be testable with additional data.

\subsection{AS 205A \label{sec:as205}}
Fits to the combined $V^2$+SED dataset have $\chi^2 \sim 4-6$, mainly due
to the poor fit to the SED data.
The flat accretion disk model provides a better fit to the combined $V^2$+SED
dataset than the puffed-up inner disk model, but both fits are of poor
quality so it is difficult to distinguish between them.
The best-fit inner disk sizes range from 0.07 to 0.14 AU, 
and inner disk temperatures are $\sim 1900$ K.
These fitted inner disk sizes are significantly larger than the
magnetospheric truncation radius computed in \S \ref{sec:rmag}
and the upper limit on co-rotation radius calculated in \S \ref{sec:corot}
(Figure \ref{fig:skyplots}).

We suggest that the poor fits to the SED data for this source
($\chi_r^2 \gg 1$; Table \ref{tab:results}) are due to near-IR emission
from an accretion shock, which has not been included in our simple models.
As discussed in \S \ref{sec:components}, sources with very high accretion
rates may produce substantial near-IR emission from hot accretion shocks;
this hot, compact emission would lead to larger inferred sizes from the SED 
but smaller inferred sizes from the $V^2$ measurements.  This is consistent
with our results for the accretion-dominated
source AS 205A ($L_{\rm acc}/L_{\rm star} \sim 10$), the only object in 
our sample for which
model fits to the SED predict larger sizes than fits to the visibilities.

Neither the flat disk model
nor the flared, puffed-up inner disk model reproduces the far-IR emission 
well.  For the flat disk model, this discrepancy is most likely due
to disk flaring, which is ignored in the model.  While the flared disk model
fits better, the measured 3-5 $\mu$m fluxes are substantially larger than
predicted by the model.  Given the extremely high inferred
accretion rate for AS 205A, we suggest that viscous dissipation of
accretion energy may lead to disk-heating, and thus additional puffing
that is not included in the model.

%More complex models are probably necessary to fit the data for
%AS 205A. First, near-IR emission from the accretion shock,
%in addition to the emission from the inner disk, must be included when
%fitting the near-IR SED.  On larger spatial scales, 
%models including accretional 
%heating through viscous dissipation, in addition to flaring, are probably 
%necessary to fit the long-wavelength SED of AS 205A.

\subsection{PX Vul \label{sec:pxvul}}
The puffed-up inner disk model provides a good fit to the SED and
visibility data for PX Vul ($\chi_r^2 = 1.09$). 
In contrast, the flat disk model provides a relatively poor fit to the
combined dataset ($\chi_r^2 = 3.10$; Figure \ref{fig:hbc293}).
The best-fit size and temperature
for the puffed-up inner disk model are $\sim 0.32$ AU and 1500 K.
Similar to AS 205A, this source has a high mass accretion rate and
displays substantial hot excess emission from an accretion shock.
However, the ratio of accretion to stellar luminosity is only $\sim 2$
for PX Vul, and there seems to be little near-IR emission due to this
hot excess; the SED is therefore fit well by our best-fit disk model.
The magnetospheric truncation radius and the co-rotation radius 
determined for PX Vul (Table 
\ref{tab:sample2}) are substantially smaller than the measured inner disk
size.  Comparison of the IRAS photometry with our models suggests that
the outer disk may be moderately flared ($\xi < 1/10$).

\section{Discussion \label{sec:disc}}

\subsection{Emerging Properties of Inner Disks around T Tauri Stars}
Inner sizes and temperatures of circumstellar disks around young stars 
have traditionally been determined by fitting  disk
models to SEDs \citep[e.g.,][]{BECKWITH+90,HILLENBRAND+92,DALESSIO+99}.  
However, recent interferometric 
observations of high-mass T Tauri and Herbig Ae/Be stars have
shown that inner disks are often much larger than predicted
by these SED models \citep{MM02,EISNER+04}.  Our new results
presented in \S \ref{sec:results} confirm this trend for 
lower-mass T Tauri stars.  

For AS 207A, V2508 Oph, and PX Vul, 
simple flat accretion disk models
suggest much smaller sizes (when fit to SEDs) than those determined
interferometrically.  
Models incorporating puffed-up inner walls and flared outer disks
provide better fits to our $V^2$ and SED data than the simple
flat disk models.  This is consistent with previous studies of more massive
Herbig Ae stars \citep{EISNER+04,LEINERT+04}, and suggests that truncated disks
with puffed-up inner walls describe lower-mass T Tauri stars in addition
to more massive objects.

The one source in our sample for which the observed $V^2$ and SED values
may be consistent with a simple flat accretion disk model is AS 205A, the
object with the highest ratio of accretion to stellar luminosity.
Recent observations of another accretion-dominated source, FU Ori,
have shown a flat disk model to be appropriate \citep{MALBET+05}.  Thus, 
the vertical structure of the inner disk may depend on the relative
magnitude of stellar and accretion luminosities. 
However, as discussed in \S \ref{sec:as205}, 
a more complicated model that accounts
for near-IR emission from accretion shocks is probably necessary to
accurately fit the data for AS 205A, and we cannot rule out
a puffed-up inner disk with our current analysis.

%Previous observations
%found that some Herbig Be stars, and the outburst source FU Ori, are also
%consistent with geometrically flat inner disk models
%\citep{EISNER+04,MALBET+05}.  All of these sources have high 
%luminosities, suggesting that high-mass or accretion-dominated
%sources may be described well by the simple flat disk model.

\subsection{Dust Sublimation and Magnetospheric Truncation \label{sec:drmag}}

The truncated disks around T Tauri and Herbig Ae stars may be explained by
dust sublimation, which depends on the disk 
temperature and dust grain properties.  
An alternative truncation mechanism is magnetospheric disruption of the disk,
which is expected to yield a range of 
inner disk truncation radii and temperatures depending on accretion rates 
and stellar 
magnetic fields \citep[e.g.,][]{KYH96}.  In reality, both mechanisms may be
operative in T Tauri disks: optically-thick dust disks (which produce most of
the observed near-IR emission) may be truncated
by dust sublimation, while an optically-thin ionized gas component may
be truncated closer to the star by the stellar magnetic field.

The calculated magnetospheric radii (\S \ref{sec:rmag}; 
Table \ref{tab:sample2}) are smaller than the 
puffed-up inner dust disk radii measured from the visibilities and SEDs 
(Table \ref{tab:results}) for all sources in our sample, suggesting
that magnetospheric truncation is not a viable mechanism for truncating
the dust disks in our sample.  Stronger magnetic fields
are an unlikely way to reconcile these differences, especially for
AS 205A and PX Vul, where $|\vec{B}_{\ast}| > 20$ kG would be required to bring
the magnetospheric truncation radii into agreement with the measured sizes.
%One potential way to reconcile these differences is stronger magnetic
%fields. We use Equation \ref{eq:rmag} to estimate the stellar magnetic field 
%strengths, $\vec{B}_{\ast}$, required so that $R_{\rm mag}=R_{\rm in}$.  
%Values for $\vec{B}_{\ast}$ are between $\sim 2-5$ kG for AS 207A and
%V2508 Oph, somewhat larger than previous measurements of T Tauri star
%magnetic field strengths \citep{JVG03}.  For AS 205A and PX Vul,
%very large stellar magnetic fields, $\vec{B}_{\ast} > 20$ kG, 
%are required to bring
%the magnetospheric truncation radii into agreement with the measured sizes.
%These high values of $\vec{B}_{\ast}$ make stronger magnetic fields an
%unlikely way to resolve the difference between measured sizes and
%magnetospheric radii.
Assuming that accreting disk material travels to $R_{\rm mag}$ in the midplane 
before being funneled along magnetic field lines onto the star, the
fact that $R_{\rm in} > R_{\rm mag}$ for all sources
suggests that the gaseous component of these
disks extends further inward 
than the dust.  

\begin{figure}
\plotone{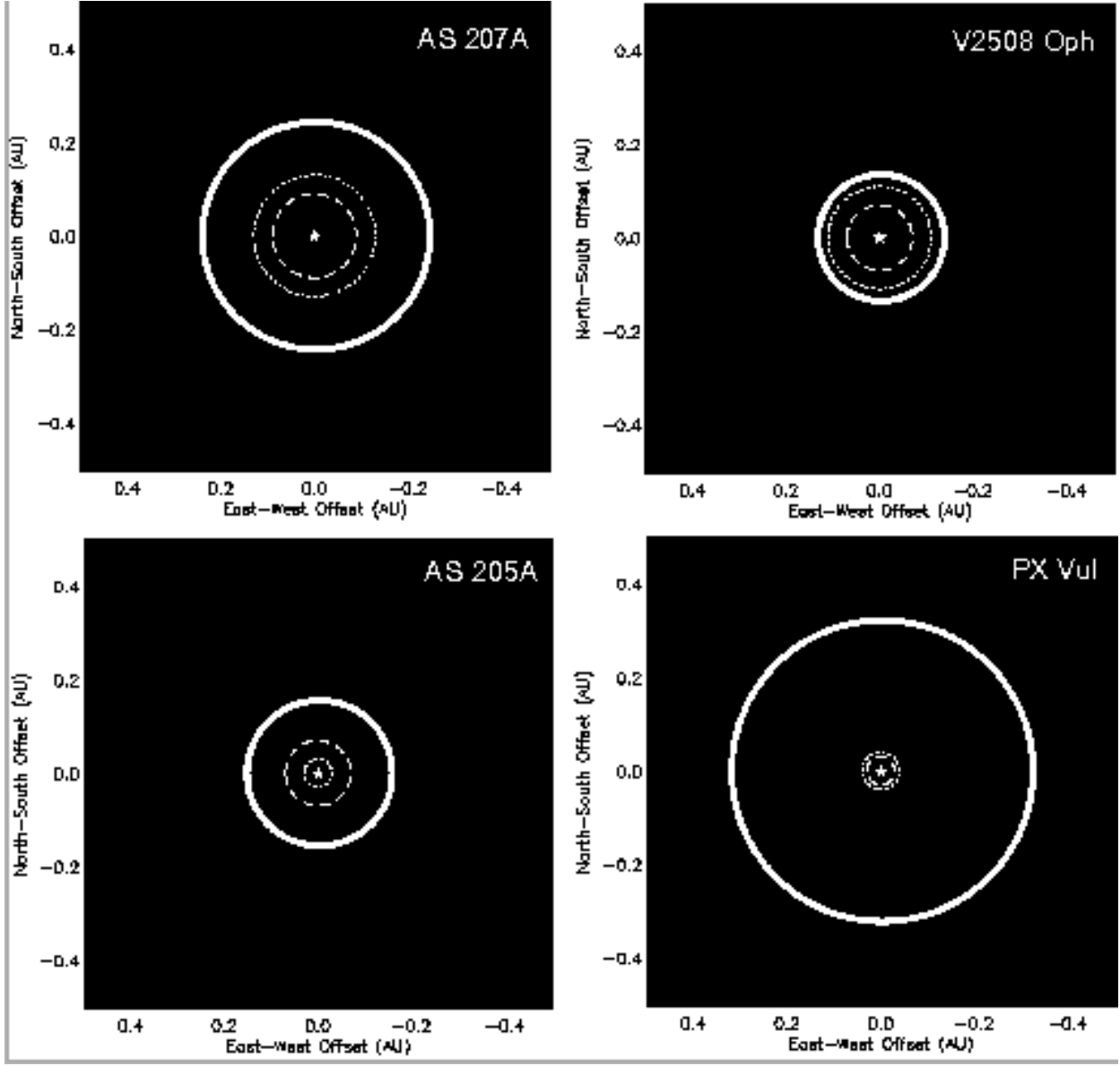}
\caption{Images of the best-fit puffed-up inner disk models for our sample,
with inner sizes and temperatures given in Table \ref{tab:results}.  Because
the puffed-up inner rims dominate the 2.2 $\mu$m emission, the images appear
ring-like.
The magnetospheric truncation radii and (limits on) co-rotation radii
(Table \ref{tab:sample2}) are indicated by dotted and dashed lines, 
respectively.  For AS 205A, V2508 Oph, and PX Vul, 
the plotted co-rotation radii
are upper limits.  The positions of the central stars are indicated with
symbols.
\label{fig:skyplots}}
\end{figure}

We speculate that dust disks are truncated
by sublimation while gaseous material in the
disk midplane extends all the way to $R_{\rm mag}$.  The smaller 
discrepancies between $R_{\rm in}$ and $R_{\rm mag}$  for 
sources with lower accretion rates (Figure \ref{fig:skyplots})
are consistent with this scenario: accretional heating 
pushes the sublimation radius outward (Equation \ref{eq:rpuff}), leading to
a larger measured inner dust disk size, while
increased pressure from accreting material compresses the magnetospheric radius
(Equation \ref{eq:rmag}).  For AS 207A and V2508 Oph, smaller 
accretion rates lead to magnetospheric truncation closer to
the sublimation radii, consistent with the data.  
In contrast, the high accretion rates in AS 205A and PX Vul may
lead to large dust sublimation radii and small magnetospheric radii, which
could explain the larger differences between $R_{\rm in}$ and $R_{\rm mag}$ in
these sources.

%The sources with smaller discrepancies between  measured dust disk sizes 
%and magnetospheric radii, AS 207A and V2508 Oph are also the youngest objects 
%in our sample (Figure \ref{fig:hr}).  
%While this may be purely coincidental,
%a larger sample of T Tauri stars, spanning a range of accretion rates and
%stellar ages, is necessary to 
%confirm our suggestion that higher accretion rates lead to larger 
%discrepancies between measured dust disk sizes 
%and magnetospheric radii.

Under standard models of magnetospheric accretion, it is expected that 
$R_{\rm mag} \la R_{\rm corot}$, since outside of co-rotation, the centrifugal
barrier would prevent accretion of material above the disk midplane 
\citep{GL79a,GL79b,KYH96,SHU+97}.  Moreover, the slow rotational velocities
of T Tauri stars \citep[compared to expectations for the collapse of
rotating clouds; e.g.,][]{HS89} are often explained by magnetic 
locking of the stellar rotation to the inner disk, which requires
$R_{\rm mag} \approx R_{\rm corot}$ \citep[e.g.,][]
{KONIGL91,SHU+94}. 
Figure \ref{fig:skyplots} shows that our results are compatible with
these models: $R_{\rm mag} \approx R_{\rm corot}$ for our sample.
Although for V2508 Oph and AS 207A, the calculated (upper limits) 
on co-rotation radii are somewhat smaller than $R_{\rm mag}$, the estimates 
agree within adopted uncertainties.  Thus, our results
are consistent with magnetospheric truncation of the gaseous component of
circumstellar disks, and magnetic 
locking of the stellar rotation and the inner (gaseous) disk edge.

\subsection{Implications for Planet Formation}
Our results indicate that dust disks
around T Tauri stars are truncated within $0.1-0.3$ AU of the central stars.
Since dust particles provide the building blocks for planetesimals, and
ultimately planets, planet formation in these systems is
unlikely interior
to $\sim 0.1$ AU.  However, our observations indicate that there {\it is}
dust in the terrestrial planet forming region (i.e., $\la 1$ AU).  
While our best-fit flat inner
disk models predict temperatures near 1 AU between 280 and 500 K
(Equation \ref{eq:racc1}), too hot
for ice condensation \citep[e.g.,][]{HAYASHI81},  the puffed-up inner
disk edges indicated by our data may cast a shadow over inner disk regions
\citep[e.g.,][]{DDN01}, leading to lower temperatures.  Thus, the 
snowline may be located at smaller radii than predicted by flat inner
disk models \citep[e.g.,][]{HAYASHI81,SL00}.
The location of the snowline has
profound implications for the formation of planets, and snowlines at smaller
radii may increase the efficiency of Earth-like planet formation 
\citep[e.g.,][]{RQL04}.

Inner disk truncation provides a natural mechanism for halting planetary 
migration
\citep[e.g.,][]{LBR96}, and may therefore be linked with the observed period 
distribution of close-in extra-solar planets.  Specifically, one expects 
migration to cease in a 2:1 resonance with the inner disk radius, 
corresponding to $0.63 R_{\rm in}$ \citep{LBR96}.  
\citet{KL02} argue that the disk
density may drop precipitously within the dust sublimation radius, and that
it is therefore the dust truncation 
radius that is important for halting migration.
For the measured inner dust radii of our sources, migrating planets would be 
halted between 0.08 and 0.20 AU.  While some extra-solar
planets are found at these radii, there is a relative dearth of planets
between $\sim 0.06-0.6$ AU, and most close-in planets are ``piled-up''
near $0.03-0.04$ AU \citep[e.g.,][]{UMS03}.  
Thus, our measured inner dust disk sizes are larger than expected based on the 
exo-planet period distribution.

This discrepancy suggests that the gaseous components of disks extend 
further toward the star than
the dust components, and that planetary migration halts in resonances with
these gaseous truncation radii as argued by \citet{LBR96}.
Gaseous material within the dust disk truncation radius is also 
consistent with the discrepancy between measured
sizes and calculated magnetospheric radii discussed above.
Assuming our inferred magnetospheric radii correspond to the inner
edges of gaseous disks, we predict resonant orbits between 0.02
and 0.08 AU from the central star, compatible with the observed
pile-up location for migrating exo-planets.
%However, the range of inner disk temperatures
%argues against sublimation as the dust truncation mechanism, suggesting
%that the gas and the dust are both truncated magnetically.

An alternate explanation for the apparent discrepancy between resonant orbits
predicted by the dust truncation sizes and those actually observed 
is that the observed exo-planet period distribution
is due to migration that occurred in an earlier evolutionary stage, when 
smaller disk truncation radii led to smaller resonant orbits.
Observations of larger samples, spanning a range of inner radii, 
are necessary to address this issue properly.

\section{Conclusions \label{sec:conc}}
We have observed three 1 M$_{\odot}$ T Tauri stars and one 2M$_{\odot}$
T Tauri star
with the Keck Interferometer to constrain
the structure of the innermost regions of their circumstellar disks.
High-resolution near-IR adaptive optics images, 
optical photometry, and optical spectra
aided in the analysis of the interferometry data,
and enabled us to estimate various properties of the systems, including
mass accretion rates and co-rotation radii.

The main result of our analysis is that inner disks around solar-mass T Tauri
stars appear similar to those around higher-mass T Tauri and Herbig Ae stars.
Specifically, the observations for most sources 
are more consistent with puffed-up inner
disk models than with geometrically flat accretion disk models.

We tested the theory of magnetospheric accretion by comparing our measured
inner dust disk radii with calculated co-rotation and magnetospheric truncation
radii.  All measured sizes are larger than the magnetospheric and co-rotation
radii.  Moreover, the difference between measured sizes and 
inferred magnetospheric/co-rotation radii seems to increase with 
accretion rate: the discrepancy is small for AS 207A and V2508 Oph, 
but large for AS 205A and PX Vul.  We suggest that accretional heating leads to
dust sublimation at radii larger than $R_{\rm mag}$.  Since higher
accretion rates cause larger sublimation radii but smaller
magnetospheric radii, this hypothesis can explain our results.
Thus, gaseous disks may extend inward to $R_{\rm mag}$, while dust disks 
are truncated further out by sublimation.  
%We note that AS 207A and V2508
%Oph are also the youngest objects in our sample, and thus the discrepancy
%between measured sizes and magnetospheric radii may depend on evolutionary
%state.  Observations of larger samples are needed to investigate these
%issues further.

Comparison of the observed inner disk sizes with the period distribution of 
extra-solar planets provides support for the hypothesis that gaseous
disks extend further inward than dust disk truncation radii, since our
measured inner disk sizes predict 2:1 resonances (which could halt migration)
farther from the star than observed for extra-solar planets.
In contrast, inferred magnetospheric radii predict resonant orbits that
are compatible with the observed semi-major axis distribution of
exo-planets.
%An alternative explanation of this disagreement is that
%planetary migration occurred in an earlier evolutionary stage when
%dust disks extended further into the star than they do at present.

\medskip
\noindent {\bf Acknowledgments:} 
The near-IR interferometry data presented in this paper were obtained
with the Keck Interferometer (KI), which is supported by NASA.  We wish
to thank the entire KI team for making these observations possible.
Observations were carried out at the W.M. Keck Observatory, which is 
operated as a scientific partnership among California Institute of Technology,
the University of California, and NASA.  The Observatory was made possible
by the generous financial support of the W.M. Keck Foundation.
The authors wish to recognize and acknowledge the cultural role and reverence 
that the summit of Mauna Kea has always had within the indigenous Hawaiian 
community. We are most fortunate to have the opportunity to conduct 
observations from this mountain.  
This publication makes use of data products from the Two Micron
All Sky Survey, which is a joint project of the University of Massachusetts
and the Infrared Processing and Analysis Center, funded by the National
Aeronautics and Space Administration and the National Science Foundation.
2MASS science data and information services were provided by the Infrared
Science Archive at IPAC.
This work has used software from the Michelson Science Center at
the California Institute of Technology. J.A.E. acknowledges support from 
a Michelson Graduate Research Fellowship.

\appendix
\section{The Spectroscopic Binary AS 205B \label{sec:as205b}}
\begin{figure}
\plotone{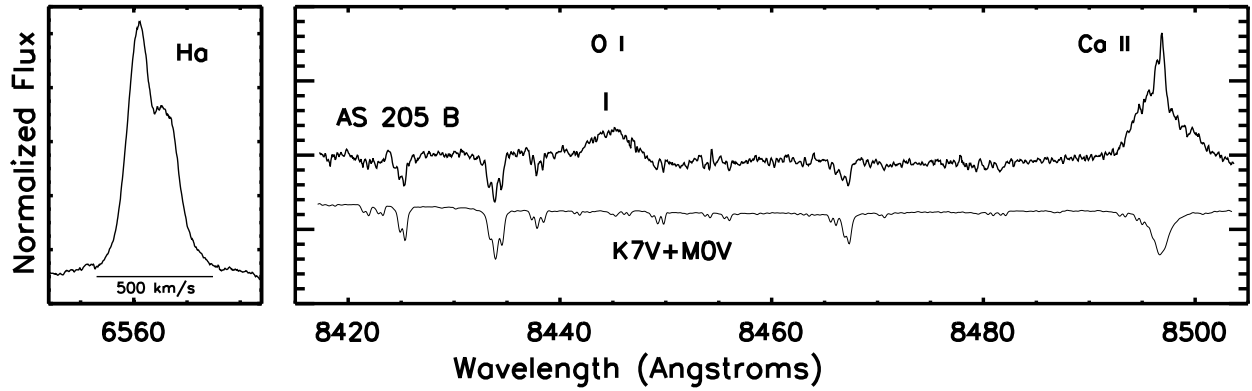}
\caption{Portions of the Keck/HIRES spectrum of the double-lined
spectroscopic binary AS 205B at H$\alpha$ and within the $I$-band; 
both panels have the same wavelength scale.  The
strong, broad H$\alpha$, OI 8446 \AA, and Ca II 8496 \AA $\:$ emission 
features suggest
on-going accretion.  The photospheric features are best matched by
combining K7 and M0 dwarfs plus an optical excess 
(see Appendix \ref{sec:as205b}).
\label{fig:as205b}}
\end{figure}

The Keck/HIRES spectrum of AS 205B 
(Figure \ref{fig:as205b}; \S \ref{sec:highres})
showed it to be a double-lined spectroscopic binary, making AS 205 a 
hierarchical triple. Since the components of AS 205B, labeled Ba and Bb, 
are of similar brightness, slowly rotating, and reasonably well separated in 
velocity, their properties can be determined somewhat independently.
We use a technique similar to that described in \S \ref{sec:stellar}, but 
tailored to fit the components of a spectroscopic binary.

Radial velocities, rotational velocities, spectral types, and continuum 
excesses for the components of AS 205B are determined from the optical
spectra.  Radial velocities and $v \sin i$
values are estimated by fitting the two peaks
of the cross-correlation function.  AS 205Ba and Bb have radial velocities
of $-0.30 \pm 0.46$ km s$^{-1}$ and $-17.71 \pm 1.11$ km
s$^{-1}$, respectively.  Both
have projected rotational velocities below our measurement limits
($v \sin i_{Ba} \leq 5.9$ km s$^{-1}$, $v \sin i_{Bb} \leq 10.2$ km s$^{-1}$); 
the larger
limit for AS 205Bb is a consequence of its fainter features.
The flux ratio of the components, their spectral types, and the continuum
excess of the system (defined here as $r = F_{excess}/(F_{Ba}+F_{Bb})$) are 
determined simultaneously by comparisons with synthetic spectroscopic binaries,
generated by combining dwarf standards at the appropriate velocities.  The
best fit is determined by minimizing the root-mean-squared difference between
AS 205B and the synthetic binary spectra over several temperature-sensitive
regions.  The components have spectral types of K7$\pm$1 (AS 205Ba) and
M0$\pm$1 (AS 205Bb), and AS 205Ba is slightly brighter
($[F_{Ba}/F_{Bb}]_{R} = 1.52\pm0.12$, $[F_{Ba}/F_{Bb}]_{I} =
1.38\pm0.11$).
The optical veiling of the system ($r_{R} = 0.83\pm0.12$, $r_{I} = 0.60\pm
0.09$) suggests on-going accretion, although it is not
possible to determine if this is onto the primary, the secondary or both.
A high accretion rate is also consistent with the strong and broad
H$\alpha$ emission (EW$ = -42.6$ \AA; 10\% width = 384 km s$^{-1}$) .  

We estimate the visual extinction and luminosity of each component using
the relative fluxes of the spectroscopic
binary components at $R$ and $I$ in combination with spectral types,
veiling, and the total AS 205B fluxes determined in \S \ref{sec:stellar}. 
We determine visual extinctions of 3.9 mag and 3.4 mag, and luminosities of
0.44 L$_\odot$ and 0.26 L$_\odot$ for AS 205Ba and Bb, respectively.
Comparison of the effective temperatures (4000 K, 3800 K) and luminosities
with the \citet{SDF00} pre-main sequence evolutionary models yields 
stellar masses of 0.74 M$_\odot$ and 0.54 M$_\odot$ and ages of 5.1 Myr and 
5.4 Myr for AS 205Ba and Bb, respectively.  
Given the uncertainties, all three components
of the AS 205 system appear to be coeval.

\epsscale{0.5}

\epsscale{1.0}

%\epsscale{1.0}
%\begin{figure}
%\plotone{figs/sdf00_eisner.epsi}
%\caption{A Hertzsprung-Russel (HR) diagram for our sample, along with
%predicted pre-main sequence evolutionary tracks from \citet{SDF00}.
%Effective temperatures are uncertain by $\sim 100$ K, and
%we estimate that the inferred
%stellar luminosities are uncertain by $\sim 30\%$.
%\label{fig:hr}}
%\end{figure}

%\begin{figure}
%\plottwo{figs/HBC254_contours.epsi}{figs/HBC264_contours.epsi}
%\caption{Adaptive optics images of the two close binaries in our sample,
%AS 205A and AS 207A, obtained with the PHARO camera on the Palomar 200-inch 
%telescope.  The contour levels are 2.5\%, 5\%, 7.5\%, 10\%, 20\%,...,
%100\% of the maximum flux in each image. 
%\label{fig:binaries}}
%\end{figure}

\clearpage
\begin{deluxetable}{lccccccccccc}
\tabletypesize{\scriptsize}
%\tabletypesize{\tiny}
\rotate
\tablewidth{0pt}
\tablecaption{Observed Properties of Sample \label{tab:sample}}
\tablehead{\colhead{Source} & \colhead{Alt. Name} & \colhead{$\alpha$} 
& \colhead{$\delta$} & \colhead{$d$} & \colhead{SpTy} & \colhead{$r_{R}$}  
& \colhead{$r_{I}$} & \colhead{$v \sin i$} & \colhead{$v_{\rm helio}$} &
\multicolumn{2}{c}{H$\alpha$} \\
 & & (J2000) & (J2000) & (pc) & & & & (km s$^{-1}$) & (km s$^{-1}$) &
EW & 10\% width}
\startdata
AS 207A & SR 9 & 16 27 40.28 & -24 22 04.0 & 160$^a$ & K5$\pm 1$ & 
$0.04 \pm 0.07$ & $-0.07 \pm 0.15$ & $15.2 \pm 0.9$ & $-7.17 \pm 0.25$ &
-16.4 & 342 \\
V2508 Oph & Oph 6 & 16 48 45.62 & -14 16 35.9 & 160$^a$ & K6$\pm 1$ & 
$0.27 \pm 0.07$ & $0.09 \pm 0.10$ & $22.9 \pm 1.0$ & $-8.26 \pm 0.56$ &
-32.5 & 278\\
AS 205A & V866 Sco & 16 11 31.40 & -18 38 24.5 & 160$^a$ & 
K5$\pm 1$ & $2.94 \pm 0.32$ & $1.95 \pm 0.20$ & $14.9 \pm 1.8$ & 
$-11.60 \pm 0.87$ & -99.6 & 388 \\
PX Vul & LkH$\alpha$ 483-41 & 19 26 40.30 & +23 53 49.0 & 420$^b$ & 
F3$\pm 2 ^{c,d}$ & 
$0.82 \pm 0.39$ & $0.62 \pm 0.45$ & $78 \pm 11$ & $-7.0 \pm 2.5$ & 
-9.4 & 512 \\ 
\enddata
\tablerefs{Spectral types, veilings (ratios of excess to 
stellar flux) at $R$ and $I$ bands, $v \sin i$ values, 
heliocentric radial velocities, H$\alpha$ equivalent widths, and
H$\alpha$ widths at 10\% of the peak, are 
determined from high-resolution optical spectroscopy
(\S \ref{sec:highres}).  
$a$--\citet{CHINI81}; $b$--\citet{HERBST+82}; $c$--
\citet{MORA+01}; $d$--\citet{HERNANDEZ+04}.  Distance estimates
are likely uncertain by 10-20\%.
}
\end{deluxetable}

\clearpage
\begin{deluxetable}{lcccccccc}
\tabletypesize{\scriptsize}
%\rotate
\tablewidth{0pt}
\tablecaption{Photometry of Observed Sources \label{tab:phot}}
\tablehead{\colhead{Source} & \colhead{$U$} & \colhead{$B$} 
& \colhead{$V$} & \colhead{$R$} & \colhead{$I$}
& \colhead{$J$} & \colhead{$H$} 
& \colhead{$K$}}
\startdata
AS 207A$^a$ & $13.52 \pm 0.53$ & $13.00 \pm 0.07$ & $11.78 \pm 0.09$ & $10.95 \pm 0.13$ & $9.80 \pm 0.09$ & $8.69 \pm 0.05$ & $7.96 \pm 0.06$ & $7.31 \pm 0.06$ \\
V2508 Oph & $15.74 \pm 1.35$ & $15.04 \pm 0.06$ & $13.45 \pm 0.03$ & $12.29 \pm 0.04$ & $10.74 \pm 0.08$ & $8.75 \pm 0.04$ & $7.73 \pm 0.07$ & $7.04 \pm 0.08$ \\
AS 205A$^{a}$ & $13.69 \pm 0.47$ & $13.74 \pm 0.06$ & $12.76 \pm 0.03$ & $11.82 \pm 0.04$ & $10.52 \pm 0.08$ & $8.63 \pm 0.13$ & $7.41 \pm 0.07$ & $6.36 \pm 0.08$ \\
PX Vul & $12.42 \pm 0.38$ & $12.35 \pm 0.05$ & $11.55 \pm 0.02$ & $11.01 \pm 0.09$ & $10.32 \pm 0.06$ & $9.33 \pm 0.14$ & $8.59 \pm 0.09$ & $7.74 \pm 0.10$ \\
%\hline
%AS 205B & $16.19 \pm 0.48$ & $16.24 \pm 0.12$ & $15.26 \pm 0.10$ & 
%$13.81 \pm 0.11$ & $12.07 \pm 0.13$ & $9.73 \pm 0.25$ & $8.35 \pm 0.12$ &
%$7.27 \pm 0.14$ \\
%AS 207B & $18.52 \pm 0.54$ & $16.81 \pm 0.12$ & $15.03 \pm 0.13$ &
%$13.84 \pm 0.16$ & $11.93 \pm 0.13$ & $10.28 \pm 0.09$ & $9.76 \pm 0.11$ &
%$9.50 \pm 0.11$ \\
\enddata
\tablerefs{$a$--For AS 207A and AS 205A, which have known companions, 
the $UBVRI$
photometry contains contributions from both components, while the $JHK$
photometry reflects only the emission from the primaries.}
\end{deluxetable}

%\clearpage
\begin{deluxetable}{lccccc}
\tabletypesize{\scriptsize}
%\rotate
\tablewidth{0pt}
\tablecaption{Binaries \label{tab:binaries}}
\tablehead{\colhead{Source} & \colhead{$Sep$} & \colhead{$P.A.$} 
& \colhead{$\Delta J$} & \colhead{$\Delta H$} & \colhead{$\Delta K$}
\\
 & ($''$) & ($^{\circ}$) & (mag) & (mag) & (mag)}
\startdata
AS 207B & 0.63 & 354 & $1.59 \pm 0.07$ & $1.79 \pm 0.09$ & $2.19 \pm 0.09$ \\
AS 205B$^a$ & 
1.31 & 213 & $1.10 \pm 0.21$ & $0.94 \pm 0.10$ & $0.91 \pm 0.12$ \\ 
\enddata
\tablerefs{$a$--AS 205B is a spectroscopic binary. See Appendix 
\ref{sec:as205b} for details.}
\end{deluxetable}

\clearpage
\begin{deluxetable}{lcccccccccccc}
\tabletypesize{\footnotesize}
%\rotate
\tablewidth{0pt}
\tablecaption{Inferred Stellar and Accretion Properties \label{tab:sample2}}
\tablehead{\colhead{Source} & \colhead{$T_{\ast}$} & \colhead{$L_{\ast}$} 
& \colhead{$R_{\ast}$} & \colhead{$A_v$} & \colhead{$M_{\ast}$} &
\colhead{$\tau_{\ast}$} & \colhead{$L_{\rm acc}$} & \colhead{$\dot{M}$} & 
\colhead{$R_{\rm mag}$} & \colhead{$R_{\rm corot}$} &
\colhead{$F_{\ast, K}$} & \colhead{$F_{{\rm D}, K}$}\\
 & (K) & (L$_{\odot}$) & (R$_{\odot}$) & (mag) & (M$_{\odot}$) & (Myr) & 
(L$_{\odot}$) & (M$_{\odot}$ yr$^{-1}$) & (AU) & (AU) & (Jy) & (Jy)}
\startdata
AS 207A & 4400 & 2.7 & 2.9 & 1.6 & 1.2 & 1.1 & 0.4 & 
$3.2 \times 10^{-8}$ & 0.13 & 0.09 & 0.59 & 0.27 \\
V2508 Oph & 4200 & 3.3 & 3.5 & 3.5 & 0.9 & 0.6 & 
1.6 & $2.3\times 10^{-7}$ & 0.11 & $\le 0.07$ & 0.82 & 0.45 \\
AS 205A & 4400 & 1.3 & 2.0 & 3.6 & 1.2 & 3.2 & 13.0 & 
$7.2 \times 10^{-7}$ & 0.03 & $\le 0.07$ & 0.27 & 2.12 \\
PX Vul & 6600 & 13.7 & 2.9 & 2.0 & 1.9 & 6.9 & 25.0 &
$1.3 \times 10^{-6}$ & 0.04 & $\le 0.03$ & 0.13 & 0.46 \\ 
\enddata
\tablerefs{Stellar parameters, accretion luminosities and rates, and
magnetospheric and co-rotation radii determined using high-resolution optical 
spectra and $UBVRI$ photometry (\S \ref{sec:systems}).  Stellar and
disk fluxes at 2.2 $\mu$m ($F_{\ast,K}$, $F_{{\rm D},K}$) determined
using Kurucz models (with measured $T_{\ast}$, $R_{\ast}$, and
adopted distances from Table \ref{tab:sample})  and
de-reddened observed photometry (\S \ref{sec:components}).
As discussed in \S \ref{sec:systems}, we estimate that $T_{\ast}$
is uncertain by $\pm 100$ K, $L_{\ast}$, $R_{\ast}$, $A_v$,
$R_{\rm mag}$, and $R_{\rm corot}$ are
uncertain by $\sim 30\%$, $L_{\rm acc}$ and $\dot{M}$ are 
uncertain by a factor of 2-3, and  $F_{\ast, K}$ and
$F_{{\rm D}, K}$ have error bars of $\sim 30-50\%$.
The uncertainties on $M_{\ast}$ and $\tau_{\ast}$ are more difficult to
ascertain since they depend on pre-main sequence evolutionary models;
however, we estimate that the relative uncertainties for these
parameters are $\sim 30-50\%$.
}
\end{deluxetable}

\clearpage
\begin{deluxetable}{lccc||ccc}
\tablewidth{0pt}
\tablecaption{Disk Parameters from Near-IR Interferometry and SEDs
\label{tab:results}}
\tablecolumns{7}
\tablehead{\colhead{ } & \multicolumn{3}{c}{Flat Disks} &
\multicolumn{3}{c}{Flared, Puffed-Up Disks} \\
\colhead{Source} & \colhead{$\chi_r^2$} & \colhead{$R_{\rm in}$} & 
\colhead{$T_{\rm in}$} & \colhead{$\chi_r^2$} & 
\colhead{$R_{\rm in}$} & \colhead{$T_{\rm in}$} \\
& & \colhead{(AU)} & \colhead{(K)} & & \colhead{(AU)} & \colhead{(K)}}
\startdata
\multicolumn{7}{c}{Combined $V^2$+SED Results} \\
\hline
AS 207A & 1.070 & $0.04_{-0.02}^{+0.01}$ & $1500_{-100}^{+300}$ & 0.208 & $0.23_{-0.10}^{+0.11}$ & $1000_{-100}^{+200}$\\
V2508 Oph & 1.409 & $0.02_{-0.01}^{+0.17}$ & $2400_{-1300}^{+100}$ & 0.981 & $0.12_{-0.10}^{+0.10}$ & $1500_{-300}^{+900}$\\
AS 205A & 3.729 & $0.07_{-0.01}^{+0.01}$ & $1900_{-100}^{+100}$ & 6.072 & $0.14_{-0.01}^{+0.01}$ & $1900_{-100}^{+100}$\\
PX Vul & 3.049 & $0.22_{-0.03}^{+0.01}$ & $1400_{-100}^{+100}$ & 1.072 & $0.32_{-0.04}^{+0.01}$ & $1500_{-100}^{+100}$\\
\hline \hline
\multicolumn{7}{c}{$V^2$-only Results} \\
\hline
AS 207A & 0.010 & $0.17_{-0.05}^{+0.04}$ & $1500$ & 0.010 & $0.25_{-0.07}^{+0.06}$ & $1000$ \\
V2508 Oph & 0.003 & $0.10_{-0.03}^{+0.04}$ & $2400$ & 0.003 & $0.20_{-0.13}^{+0.04}$ & $1500$ \\
AS 205A & 0.211 & $0.07_{-0.01}^{+0.01}$ & $1900$ & 0.224 & $0.13_{-0.01}^{+0.01}$ & $1900$ \\
PX Vul & 0.234 & $0.23_{-0.01}^{+0.01}$ & $1400$ & 0.234 & $0.34_{-0.02}^{+0.03}$ & $1500$ \\
\hline \hline
\multicolumn{7}{c}{SED-only Results} \\
\hline
AS 207A & 0.473 & $0.04_{-0.02}^{+0.01}$ & $1500_{-100}^{+300}$ & 0.327 & $0.23_{-0.23}^{+0.11}$ & $1000_{-100}^{+300}$\\
V2508 Oph & 0.969 & $0.02_{-0.01}^{+0.03}$ & $2400_{-800}^{+100}$ & 0.870 & $0.07_{-0.07}^{+0.08}$ & $1900_{-500}^{+600}$\\
AS 205A & 6.851 & $0.06_{-0.01}^{+0.01}$ & $2000_{-100}^{+200}$ & 4.796 & $0.23_{-0.03}^{+0.01}$ & $1600_{-100}^{+100}$\\
PX Vul & 1.341 & $0.05_{-0.01}^{+0.03}$ & $2400_{-400}^{+100}$ & 0.842 & $0.21_{-0.04}^{+0.06}$ & $1800_{-200}^{+200}$\\
\hline
\enddata
\tablerefs{Columns 2-4 list the reduced $\chi^2$ values, inner radii, and
inner temperatures for best-fit flat accretion disk models.  Columns
5-7 list $\chi_r^2$, $R_{\rm in}$, and $T_{\rm in}$ for best-fit
puffed-up inner disk models.  Results are shown for fits to combined 
$V^2$+SED datasets as well as $V^2$ and SEDs individually.  For the $V^2$-only
fits, we assumed the best-fit temperature from the combined $V^2$+SED
analysis, fitting only for $R_{\rm in}$.}
\end{deluxetable}

\clearpage
\begin{deluxetable}{lccc||ccc}
\tablewidth{0pt}
\tablecaption{Measured versus Predicted Inner Disk Sizes
\label{tab:predictions}}
\tablecolumns{7}
\tablehead{\colhead{ } & \multicolumn{3}{c}{Flat Disks} &
\multicolumn{3}{c}{Flared, Puffed-Up Disks} \\
\colhead{Source} & \colhead{$R_{\rm in,meas}$} & 
\colhead{$R_{\rm in, \dot{M}=0}$} & 
\colhead{$R_{\rm in, \dot{M} \ne 0}$} & \colhead{$R_{\rm in,meas}$} & 
\colhead{$R_{\rm in, \dot{M}=0}$} & 
\colhead{$R_{\rm in, \dot{M} \ne 0}$} \\
& \colhead{(AU)} & \colhead{(AU)} & \colhead{(AU)} & \colhead{(AU)} & 
\colhead{(AU)} & \colhead{(AU)}}
\startdata
AS 207A & 0.17$_{-0.05}^{+0.04}$ & 0.03 & 0.04 & 0.23$_{-0.10}^{+0.11}$ & 0.26 & 0.28 \\
V2508 Oph & 0.10$_{-0.03}^{+0.04}$ & 0.02 & 0.03 & 0.12$_{-0.10}^{+0.10}$ & 0.13 & 0.16 \\
AS 205A & 0.07$_{-0.01}^{+0.01}$ & 0.01 & 0.07 & 0.14$_{-0.01}^{+0.01}$ & 0.05 & 0.17 \\
PX Vul & 0.23$_{-0.01}^{+0.01}$ & 0.06 & 0.15 & 0.32$_{-0.04}^{+0.01}$ & 0.27 & 0.46 \\
\enddata
\tablerefs{Measured sizes ($R_{\rm in, meas}$) taken from Table 
\ref{tab:results}, compared to expectations for disk models based on
inferred inner disk temperatures and stellar parameters.  
For the puffed-up inner disk models, we use the sizes measured from
combined $V^2$+SED analysis, while for the flat disk models, where models
generally provide poor fits to the $V^2$ and SED values simultaneously, we use
measured sizes from $V^2$-only analysis.  Expected inner disk sizes
for the flat and puffed-up disk models 
are calculated from Equations \ref{eq:racc1}--\ref{eq:rpuff},
assuming no accretion ($R_{{\rm in}, \dot{M}=0}$) and using the 
inferred accretion rates 
from Table \ref{tab:sample} ($R_{{\rm in}, \dot{M} \ne 0}$).}
\end{deluxetable}

%\clearpage
%\begin{deluxetable}{lccc}
%\tablewidth{0pt}
%\tablecaption{Magnetospheric Radii
%\label{tab:rmag}}
%\tablehead{\colhead{Source} & \colhead{$R_{\rm mag}$} & 
%\colhead{$R_{\rm in,meas}$} & \colhead{$\vec{B}_{\ast}$}\\
% & (AU) & (AU) & (kGauss)}
%\startdata
%AS 205A & 0.03 & 0.14 & 27.3 \\
%AS 207A & 0.13 & 0.23 & 5.22 \\
%V2508 Oph & 0.12 & 0.12 & 2.01 \\
%PX Vul & 0.07 & 0.31 & 26.3 \\
%\enddata
%\tablerefs{Measured sizes are taken from Table \ref{tab:results}, using the
%results of the puffed-up inner disk models fit to combined $V^2$+SED datasets.  
%$R_{\rm mag}$ is calculated
%using Equation \ref{eq:rmag} with accretion rates from Table
%\ref{tab:sample2} and assuming a stellar magnetic field of 2 kG.
%$\vec{B}_{\ast}$ is calculated from Equation \ref{eq:rmag} assuming
%$R_{\rm mag}=R_{\rm in, meas}$.}
%\end{deluxetable}

\end{document}